\documentclass[%
reprint,
nofootinbib,
amsmath,amssymb,
aps,
pra,
floatfix,
]{revtex4-2}

\usepackage{bm}
\usepackage[utf8]{inputenc}
\usepackage[T1]{fontenc}
\usepackage{dsfont}
\usepackage{parskip}
\usepackage{graphicx}
\usepackage{xcolor}
\usepackage{quantikz}

\usepackage{svg} % to include svg images from pptx
\usepackage{hyperref}
\hypersetup{colorlinks=true,linkcolor=blue,citecolor=blue,filecolor=blue,urlcolor=blue}

\usepackage{array} % Required for defining custom column types
\usepackage{booktabs}
\usepackage{multirow}
\usepackage{dcolumn}% Align table columns on decimal point

\newcolumntype{L}{>{\raggedright\arraybackslash}p{3cm}}
\usepackage{arydshln} %for dashed lines in tables

\usepackage{float}
\usepackage{microtype}

\usepackage[super]{nth}
\usepackage{amsfonts}
\usepackage{amsmath, amsthm, amssymb}
\usepackage{nicefrac}
\usepackage{bm}% bold math
\usepackage{physics}% bold math
\usepackage{braket}
\usepackage{siunitx}

\newcommand*{\kmnistName}{\texttt{kMNIST}28}
\newcommand*{\plasticcName}{\texttt{plasticc}}
\newcommand*{\hiddenmanifoldName}{\texttt{hidden} \texttt{manifold}}

\newcommand{\KRBF}{\mathbf{K}^{\text{(RBF)}}}
\newcommand{\kRBF}{k^{\text{(RBF)}}}
\newcommand{\Kpoly}[1]{\mathbf{K}^{(#1)}}
\newcommand{\kpoly}[1]{k^{(#1)}}
\newcommand{\KFQK}{\mathbf{K}^{\text{(FQK)}}}
\newcommand{\KPQK}{\mathbf{K}^{\text{(PQK)}}}
\newcommand{\kFQK}{k^{\text{FQK}}}
\newcommand{\kPQK}{k^{\text{PQK}}}
\newcommand{\varKD}{\text{Var}_{\mathcal{D}}[\mathbf{K}]}
\newcommand{\varksep}{\text{Var}_{\mathbf{x}, \mathbf{x}^\prime}[k_{R_X}(\tilde{\mathbf{x}},\tilde{\mathbf{x}}^\prime)]}

\usepackage{lipsum}
\usepackage{comment}
\newcommand{\red}{\color{\red}}

\newcommand{\commentedout}[1]{}

\begin{document}

\title{On the similarity of bandwidth-tuned quantum kernels and classical kernels}

\author{Roberto Flórez-Ablan}
\email{roberto.florez.ablan@ipa.fraunhofer.de}
\affiliation{Fraunhofer Institute for Manufacturing Engineering and Automation IPA, Nobelstraße 12, D-70569 Stuttgart, Germany}

\author{Marco Roth}
\email{marco.roth@ipa.fraunhofer.de}
\affiliation{Fraunhofer Institute for Manufacturing Engineering and Automation IPA, Nobelstraße 12, D-70569 Stuttgart, Germany}

\author{Jan Schnabel}
%\email{jan.schnabel@ipa.fraunhofer.de}
\thanks{Corresponding author: \href{mailto:jan.schnabel@ipa.fraunhofer.de}{jan.schnabel@ipa.fraunhofer.de}}
\affiliation{Fraunhofer Institute for Manufacturing Engineering and Automation IPA, Nobelstraße 12, D-70569 Stuttgart, Germany}

\date{\today}
 
\begin{abstract}
Quantum kernels (QK) are widely used in quantum machine learning applications; yet, their potential to surpass classical machine learning methods on classical datasets remains uncertain. This limitation can be attributed to the exponential concentration phenomenon, which can impair generalization. A common strategy to alleviate this is bandwidth tuning, which involves rescaling data points in the quantum model to improve generalization. In this work, we numerically demonstrate that optimal bandwidth tuning results in QKs that closely resemble radial basis function (RBF) kernels, leading to a lack of quantum advantage over classical methods. Moreover, we reveal that the size of optimal bandwidth tuning parameters further simplifies QKs, causing them to behave like polynomial kernels, corresponding to a low-order Taylor approximation of a RBF kernel. We thoroughly investigate this for fidelity quantum kernels and projected quantum kernels using various data encoding circuits across several classification datasets. We provide numerical evidence and derive a simple analytical model that elucidates how bandwidth tuning influences key quantities in classification tasks. Overall, our findings shed light on the mechanisms that render QK methods classically tractable.
\end{abstract}

\maketitle

\section{Introduction}
Quantum kernel (QK) methods have emerged as a promising branch of quantum machine learning (QML) research~\cite{Biamonte.2017, Schuld.2021b, Schuld.2022, Schuld.2019b, Schuld.26012021, Huang.2021, Havlicek.2019, Schuld.2019b, Thanasilp.2024, Gil_Fuster_2024} with several proposals for promising applications in various domains~\cite{challengesQMLreview, Wang_2017,Peruzzo_2014,verdon2019quantumhamiltonianbasedmodelsvariational,C_rstoiu_2020, Cong_2019, Meyer_2021,Beckey_2022}. Here, a central achievement was that it has been formally shown~\cite{Schuld.2019b, Schuld.26012021} that QK methods can be embedded into the rich mathematical framework of conventional kernel theory~\cite{Scholkopf., Hastie.2009}. The key idea of QKs is that information is processed by mapping data into the exponentially large Hilbert space of quantum states. In practice this is realized by data encoding circuits~\cite{Schuld.2019b, lloyd2020, Schuld.2021, Jerbi.2023}. Eventually, the quantum computer is only required to compute the QK matrix, which is subsequently passed to a conventional kernel method such as support vector machine (SVM). There are two common approaches to evaluate the QK functions: fidelity quantum kernels (FQK)~\cite{Schuld.2019b, Havlicek.2019, Blank2020, Hubregtsen.2022} and projected quantum kernels (PQK)~\cite{Huang.2021, Gan.22112023, suzuki2023effectalternatinglayeredansatzes}. 

Although a rigorous quantum advantage of QKs has been proven for the discrete logarithm problem based on a Shor-type data encoding~\cite{Liu.2021}, an overall advantage of QKs over classical ML methods on real-world problems and more realistic data encodings has not yet been achieved~\cite{schnabel2024quantumkernelmethodsscrutiny, Bowles.11032024, Huang.2021, Thanasilp.2024, Kubler.}. In this regard,~\citet{Kubler.} provide insights into the inductive bias of QKs and point out that they may only offer an advantage over classical models in cases where encoding knowledge about the problem is easy for quantum models and hard for classical models. Early work, cf. e.g., Ref.~\cite{Havlicek.2019}, highlighted that embedding data into a Hilbert space whose dimension grows exponentially could provide a path to quantum speed-ups in ML tasks. However, recent theory shows that this exponential growth actually constitutes another central limitation of QKs, which can cause the matrix elements to exponentially concentrate around a fixed value~\cite{Thanasilp.2024}. As the number of qubits increases, one effectively computes inner products between states in an exponentially large space, which induces the vanishing similarity problem~\cite{Suzuki.29102022}. Although this preserves the convexity of the training landscape of kernel-based methods, one ends up with a trivial QK with detrimental effect on the generalization properties~\cite{gilfuster2025relationtrainabilitydequantizationvariational}. This phenomenon constitutes a similar barrier as the barren plateau problem in quantum neural networks~\cite{Thanasilp.2024,kairon2025equivalenceexponentialconcentrationquantum, Arrasmith.2022,Cerezo.2021,Cerezo.2021c,Pesah.2021,larocca2024reviewbarrenplateausvariational,McClean_2018,Ragone_2024}.

\begin{figure}
    \centering
    \includegraphics[width=\linewidth]{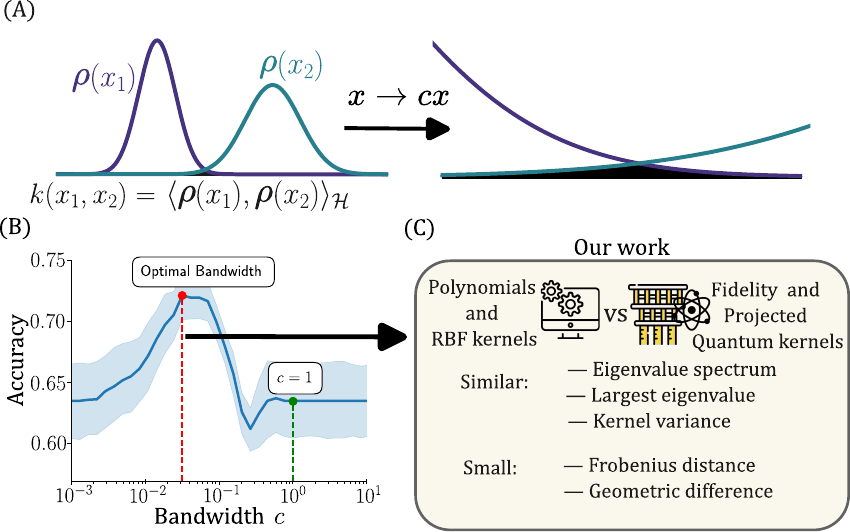}
    \caption{Schematic overview of our work. (A) Information is encoded into quantum states and QKs are generally defined via the inner products between two vectors in this exponentially large space. Thus, QKs suffer from the vanishing similarity problem due to the size of the Hilbert space. Multiplying the input data by a global bandwidth parameter $c$ artificially increases the inner product between the encoded vectors and (B) improves the performance of QKs but results in classically tractable kernels. In this work (C), we show that optimal bandwidth parameters are so small that the properties of the QKs become similar to those of CKs.}
    \label{fig:overview_diagram}
\end{figure}

To circumvent this setback several alternative QK variants have been formulated~\cite{Huang.2021, suzuki2023effectalternatinglayeredansatzes, Suzuki.29102022, glick_covariant_2024}. Other approaches are based on parametrized data encoding circuits and kernel target alignment to optimize the QK~\cite{lloyd2020, Hubregtsen.2022}. Furthermore, it has been shown that using bandwidth tuning, i.e., rescaling input data by a constant factor, controls the model's inductive bias by beneficially changing the eigenvalue spectra as well as the task-model alignment and thus enables generalization of QK models~\cite{Shaydulin.2022, Slattery.2023, canatatheorybandwidthenablesgeneralization, canatatheoryclassicalkernels, canatatheoryclassicalkernels}. However, improving the generalization of QK models using bandwidth tuning comes at the cost of losing quantum advantage~\cite{Slattery.2023}. For selected (classical) classification tasks, it has been shown that with increasing qubit count the optimal FQKs become substitutable by classical kernel (CK) models. Additionally, bandwidths that induce an appropriate inductive bias to the QK spectrum and thus enable generalization lead to small geometric differences~\cite{Egginger.2024}. Complementarily, ~\citet{Bowles.11032024} observed that QK methods have a surprisingly similar performance to SVMs with a RBF kernel across various classification tasks. 

Systematically investigating whether a classical model can replace a given quantum model is known as dequantization~\cite{dequantizing_QML_using_TN}. Here, \citet{landman2022classicallyapproximatingvariationalquantum} proposed to train a classical kernel  with carefully chosen kernel functions approximated via Random Fourier Features (RFF) to efficiently dequantize a QML model. The necessary and sufficient conditions under which RFF efficiently dequantize QML have been formally established in Ref.~\cite{Sweke_2025_potentialandlimitation}. Several highly efficient approaches to dequantization with implications to RFF-based methods, leverage techniques from tensor networks (TN)~\cite{dequantizing_QML_using_TN, sweke2025kernelbaseddequantizationvariationalqml}. A generalization to this with further avenues to dequantization is captured by the framework of entangled tensor kernels~\cite{shin2025newperspectivesquantumkernels}, which encompasses all embedding QKs. Recently, RFF-dequantization has been extended to classification and regression tasks based on both QSVMs and QNNs in Ref.~\cite{sahebi2025}.

Since kernels generally define similarity measures on the input space, it is a natural question to ask which similarity measures QKs induce in high dimensions. In this regard, it was already indicated in Refs.~\cite{Haug_2023, haug2021optimaltrainingvariationalquantum} that in the small bandwidth limit FQKs based on product state encoding circuits can be approximated by RBF kernels for vanishing bandwidths.

We contribute to comprehensively answering this question by thoroughly investigating both FQKs and PQKs across various encoding circuits from common QML literature for different classification tasks. Furthermore, we explore the mechanisms that make bandwidth-tuned QKs classically tractable. Here, we go beyond previous work and reveal how key quantities of QKs, i.e., the variance of the kernel Gram matrix, the largest eigenvalue, and the expressivity of the encoding circuits depend on the bandwidth tuning hyperparameter. We identify three distinct regimes and connect the various properties of QKs. To understand these findings we derive a simple analytic reference model that is able to capture the behavior of FQK and PQK quantities. We numerically show that all simulated QK matrices produce similarity measures that closely resemble radial basis function (RBF) kernels. Moreover, we demonstrate that the size of optimal bandwidth tuning parameters is so small that QKs simplify even further and effectively behave like polynomial kernels corresponding to a low-order Taylor approximation of a RBF kernel. In this regard, our work aligns with recent RFF- and TN-based approaches toward dequantizing QK models. While these works establish theoretical frameworks, we explicitly study dequantization as a function of bandwidth tuning for both FQKs and PQKs. Based on the three regimes, we provide a straightforward assessment of whether QKs can be replaced by common CKs rather than RFFs. Although bandwidth-tuned FQKs and PQKs are captured by the ETK formalism, studying its actual impact on classical tractability was not a particular focus therin. As we support these findings with comprehensive heuristics across various data encoding circuits and datasets, our study may serve as an accessible tool to probe the dequantization of QKs. The key parts of the paper are shown schematically in Fig.~\ref{fig:overview_diagram}. 

The remainder of this work is organized as follows. Section~\ref{sec:background} introduces the theoretical background of QK methods and bandwidth tuning as well as the key concepts to measure exponential concentration, expressivity and generalization properties. Subsequently, we first unfold our experimental setup in Sec.~\ref{sec:results} before we delve into our findings. We thoroughly analyze our simulation results, showcasing that QKs closely resemble RBF and polynomial kernels
as a function of the bandwidth tuning parameter and share important properties. Hereafter, we provide an analytical model that captures those findings and detail the behavior of optimally bandwidth-tuned QKs. In Sec.~\ref{sec:discussion}, we discuss generalizations and potential limitations, and situate our results into the current discourse. Finally, we summarize our work in Sec.~\ref{sec:conclusion}.

\section{Background\label{sec:background}}
Kernels are real-valued\footnote{In principle kernels can be complex-valued but in the context of this work it suffices to restrict ourselves to real-valued kernels.} functions $k:\mathcal{X}\times\mathcal{X}\to\mathbb{R}$ of two data points from some input domain $\mathcal{X}$. Kernel functions define an inner product of data points that are transformed into a potentially infinite-dimensional feature space $\mathcal{F}$ using a feature map $\varphi:\mathcal{X}\rightarrow\mathcal{F}$. Hence, one can think of kernels as a similarity measure between two data points. A central result from kernel theory is the representer theorem~\cite{Scholkopf., Schoellkopf2001representer}, which states that the function $f$ that solves the supervised learning problem based on a training dataset $\mathcal{D}=\lbrace (\mathbf{x}_1,y_1),\ldots,(\mathbf{x}_N, y_N)\rbrace$, with $\mathbf{x}_i\in\mathbb{R}^n$
and $y_i \in \mathbb{R}$, can always be represented as a finite linear combination of the kernel function evaluated on the training data.

\begin{figure*}
    \centering
    \includegraphics[width=0.75\linewidth]{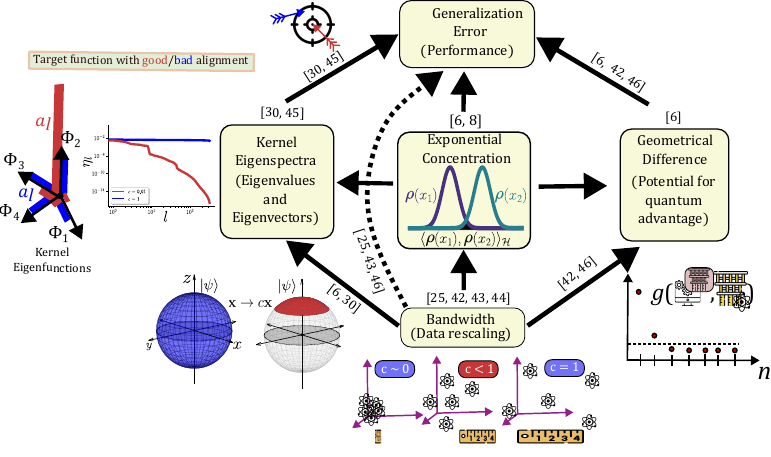}
    \caption{Interaction of various important properties and metrics for the evaluation of QKs (from \textbf{bottom} to \textbf{top}). (Left path) Bandwidth-tuning reduces the quantum circuits expressivity and changes the eigenvalue spectra of QKs (Bloch sphere/eigenvalue spectrum inspired/reproduced by Ref.~\cite{canatatheorybandwidthenablesgeneralization}). This results in QKs that can generalize. (Right path) Bandwidth-tuned QKs lose the potential for quantum advantage with increasing qubit count, as measured by the geometrical difference $g(\mathbf{K}_{\mathrm{C}},\mathbf{K}_{\mathrm{Q}})$, which indicates how much better a CK generalizes than a QK. (Central path) Bandwidth tuning mitigates exponential concentration of QKs as measured by the variance of the kernel Gram matrix entries. The exponential concentration phenomenon leads to QKs that cannot generalize. (Indirect path) Bandwidth-tuning can be seen as a technique to improve the performance of QK methods.}
    \label{fig:condensed_diagram}
\end{figure*}

This is the foundation of respective kernel methods such as SVMs~\cite{Scholkopf.} or kernel ridge regression (KRR)~\cite{Murphy.2012}. All these algorithms are based on the kernel \emph{Gram matrix} $[\mathbf{K}]_{ij}=k(\mathbf{x}_i,\mathbf{x}_j)$.

In QML, information is processed by encoding (embedding) input data $\mathbf{x}\in\mathcal{X}$ into a quantum state~\cite{Schuld.2019b}
\begin{align}
    \rho(\mathbf{x}) = U(\mathbf{x})\rho_0U(\mathbf{x})^\dagger\,,
\end{align}
using an encoding unitary $U(\mathbf{x})$ applied to the initial $n_\text{q}$-qubit state $\rho_0=\vert 0\rangle\!\langle 0\vert$. Defining this as the data encoding feature map~\cite{Schuld.26012021, Havlicek.2019}, lays the foundation for formulating quantum kernel methods as a classical kernel method whose kernel is computed using a quantum computer. In analogy to conventional kernel theory, the natural way to define a QK, relies on leveraging the native geometry of the underlying quantum state space and use the Hilbert-Schmidt inner product, i.e.,
\begin{equation}
    \label{eq:FQK}
    k^{\mathrm{FQK}}(\mathbf{x}, \mathbf{x^\prime}) = \tr[\rho(\mathbf{x})\rho(\mathbf{x^\prime})]\,.
\end{equation}
This represents a fidelity-type metric, wherefore Eq.~\eqref{eq:FQK} is commonly referred to as \emph{fidelity quantum kernel}.

Equation~\ref{eq:FQK} is an inner product between two $2^{n_\text{q}}$-dimensional objects, which constitutes a severe limitation for increasing problem size~\cite{Huang.2021, Thanasilp.2024}. \emph{Projected quantum kernels} have been proposed to mitigate this problem~\cite{Huang.2021}. They are typically defined using $k$-particle reduced density matrices ($k$-RDMs) to project the quantum states to an approximate classical representation which are subsequently used as features in a conventional kernel function. The most common definition is based on measuring the $1$-RDM on all qubits with respect to the Pauli operators $P \in \lbrace X, Y, Z\rbrace$ and hereafter use a RBF outer kernel function
\begin{equation} 
    \label{eq:PQK_RBF}
    k^{\mathrm{PQK}}(\mathbf{x},\mathbf{x}^\prime) = \exp\left(-\gamma \sum_{k, P} \lbrace\mathrm{tr}{\left[P\rho_k(\mathbf{x})\right]} - \mathrm{tr}{\left[P\rho_k(\mathbf{x}^\prime)\right]} \rbrace^2 \right)\,,
\end{equation}
where  $\gamma\in\mathbb{R}_+$ is a hyperparameter and the $1$-RDM is the partial trace of the quantum state $\rho(\mathbf{x})$ over all qubits except for the $k$-th qubit.

Under certain conditions the values of FQKs and PQKs can \emph{exponentially concentrate} with increasing number of qubits around a fixed value $\mu$~\cite{Thanasilp.2024}.
Consequently, with increasing dimension, the off-diagonal elements in the corresponding QK Gram matrices contain, with high probability, no information about the input training data. Thus, the model attains a trivial form and loses its generalization properties. An important quantity to diagnose (probabilistic) exponential concentration is to analyze the variance of respective QK Gram matrices $\mathbf{K}$ evaluated for a given training dataset $\mathcal{D}$, i.e.
\begin{equation}
    \label{eq:variance_discrete}
    \varKD = \frac{1}{N} \sum_{i \neq j} \left( k(\mathbf{x}_i, \mathbf{x}_j) - \mu \right)^2\,,
\end{equation}
where $\mathbf{x}_i,\mathbf{x}_j\in\mathcal{D}$, and $\mu$ corresponds to the mean value of all kernel entries, i.e., $\mu = \mathbb{E}_{\mathcal{D}}(\mathbf{K}) = 1/N^2\sum_{i,j}^N k(\mathbf{x}_i, \mathbf{x}_j)$.

One mechanism that can lead to exponential concentration is related to the expressivity of the underlying data encoding unitary $U(\mathbf{x})$, or the unitary ensemble $\mathbb{U}_{\mathcal{X}}=\{U(\mathbf{x}) | \, \mathbf{x} \in \mathcal{X}\}$ 
generated via $U(\mathbf{x})$ over all possible input data vectors $\mathbf{x}\in\mathcal{X}$. The expressivity $\epsilon_{\mathbb{U}_{\mathcal{X}}}$ of the ensemble quantifies how close it is from a $2$-design or, in broad terms, how close the ensemble uniformly covers the unitary group. This measure of expressivity emerged in the context of the variational quantum eigensolver, where exploring a large part of the Hilbert space is important~\cite{Sim.2019, Nakaji_2021}. Formally, it is defined as the difference of an integral over the Haar ensemble $\mathcal{V}_{\mathrm{Haar}}$ and an integral $\mathcal{V}_{\mathbb{U}}$ over $\mathbb{U}$ (cf. Appendix~\ref{appendix:expressivity} for details)
\begin{equation}
\label{eq:expressiblity_definition}
\epsilon_{\mathbb{U}_{\mathcal{X}}} = \norm{\mathcal{V}_{\mathrm{Haar}} - \mathcal{V}_{\mathbb{U}_{\mathcal{X}}}}_1 \,.
\end{equation}
For a maximally expressive ensemble, the measure yields $\epsilon_{\mathbb{U}_{\mathcal{X}}}=0$. To gain intuition about expressivity-induced concentration, consider the evaluation of FQKs, cf. Eq.~\eqref{eq:FQK}, which requires computing the inner product between two vectors in an exponentially large Hilbert space. For highly expressive data encodings, this essentially corresponds to computing the inner product between two approximately random, and thus orthogonal vectors.

An important concept which generally helps to cure exponential concentration effects is bandwidth tuning~\cite{Shaydulin.2022, canatatheoryclassicalkernels, canatatheorybandwidthenablesgeneralization}.
The key idea, as illustrated in Fig.~\ref{fig:condensed_diagram}, is to rescale the inputs by a global bandwidth-tuning parameter $c$, i.e., $\tilde{\mathbf{x}}=c \mathbf{x}$, to limit the embedding to a smaller region~\cite{canatatheorybandwidthenablesgeneralization}. This is similar to optimizing respective length scale parameters in CKs~\cite{Shaydulin.2022}. There exists an optimal bandwidth $c^*$ that significantly improves the prediction performance of quantum kernel methods (indirect path of Fig.~\ref{fig:condensed_diagram}), thus enabling generalization~\cite{canatatheoryclassicalkernels, Shaydulin.2022, Egginger.2024, Gan.22112023}.

The generalization behavior of kernels can be analyzed by inspecting the spectrum of the (normalized) Gram matrix. Given its eigendecomposition, it can be shown that the generalization error can be written as the sum of the error of each spectral mode. The key observation from this analysis is that a flat eigenvalue spectrum requires more training data than a decaying spectrum. Additionally, if the size of the largest eigenvalue $\eta_{\mathrm{max}}$ is small compared to the number of data points, the model is not able to generalize~\cite{Kubler.}. Furthermore, bandwidth tuning changes the spectrum from flat to decaying \cite{canatatheorybandwidthenablesgeneralization} and thus enables the QKs to generalize; cf. left path of Fig.~\ref{fig:condensed_diagram}.

However, inducing an appropriate inductive bias to the QK spectrum by bandwidth tuning comes at the cost of losing the potential for quantum advantage. This has been investigated in Ref.~\cite{Slattery.2023} for bandwidth-tuned FQKs on selected (classical) classification problems. The separation between quantum and classical kernelized models can be measured using the geometric difference metric~\cite{Huang.2021} 
\begin{align}
\label{eq:geo_diff_reg}
g(\mathbf{K}_1,\mathbf{K}_2) = 
 \sqrt{\left\|\sqrt{\mathbf{K}_{1}} \sqrt{\mathbf{K}_{2}} \left(\mathbf{K}_{2} + \lambda \mathds{1} \right)^{-2}\sqrt{\mathbf{K}_{2}}\sqrt{\mathbf{K}_{1}}\right\|_{\infty}}\,,
\end{align}
where $\lVert\cdot\rVert_{\infty}$ is the spectral norm. In this work, we are usually interested in the case where $\mathbf{K}_1=\mathbf{K}_{\mathrm{C}}$ and $\mathbf{K}_2=\mathbf{K}_{\mathrm{Q}}$ are classical and quantum kernel matrices with $\mathrm{tr}(\mathbf{K}_{\mathrm{C}}) = \mathrm{tr}(\mathbf{K}_{\mathrm{Q}}) = N$, respectively, and $N$ the number of training samples. The geometric difference is an asymmetric metric and can be used to quantify the potential for quantum advantage. It can be shown that only if $g(\mathbf{K}_{\mathrm{C}},\mathbf{K}_{\mathrm{Q}})\propto\sqrt{N}$ holds, the necessary condition for potential quantum advantage is fulfilled; cf. right part of Fig.~\ref{fig:condensed_diagram}.

\section{\label{sec:results}Results}

To understand what leads the bandwidth-tuned quantum models to become classically tractable and to quantify which similarity functions FQKs and PQKs effectively define, we perform numerical simulations. All simulations are based on sQUlearn~\cite{kreplin2025squlearn} with the PennyLane~\cite{Bergholm.12112018} statevector simulator device. The code to reproduce all of the simulation results of this study is available via GitHub~\cite{robertogit}.

We choose a similar setting to Refs.~\cite{Shaydulin.2022, Slattery.2023} and perform investigations on three classification datasets: \plasticcName \, \cite{team.28092018}, \kmnistName \, \cite{Clanuwat2018} (classification of digits 2 and 8) and \hiddenmanifoldName \, \cite{Bowles.11032024, Goldt_2020}. For each dataset, we randomly sample $N=400$ points and split them into 320 training and 80 test samples, respectively. Hereafter, we first employ standardization by removing the mean and apply principal component analysis (PCA) as implemented in scikit-learn~\cite{scikitPedregosa.} to reduce the dimensionality to $n$ features. To investigate the effect of bandwidth-tuning, all $N$ features $\mathbf{x}_i$ are rescaled by a global factor $c>0$ as $\tilde{\mathbf{x}}_i=c \mathbf{x}_i$. For both quantum and classical approaches, we use a SVM for solving the classification tasks. To obtain robust and statistically reliable results, each experiment is repeated six times using different seeds for randomly splitting the data into train and test subsets and we conduct a grid search for obtaining corresponding optimal regularization parameters (i.e., $C$ for the SVM) and bandwidths $c^*$. For comparing QKs to CKs, we consider classical RBF kernels as implemented in scikit-learn and set the length-scale parameter \texttt{gamma}$=1$, i.e.,
\begin{align}
    \label{eq:RBF} \left(\mathbf{K}^{\text{(RBF)}}\right)_{ij} &= k^{\text{(RBF)}}(\tilde{\mathbf{x}}_i, \tilde{\mathbf{x}}_j) =e^{-\norm{\tilde{\mathbf{x}}_i - \tilde{\mathbf{x}}_j}^2}\,,
\end{align}
as well as order $2t$ polynomial kernels corresponding to the Taylor expansion of Eq.~\eqref{eq:RBF} truncated after the $t$-th term~\cite{inria-00325810, RING2016107,cotter2011explicitapproximationsgaussiankernel}. For $t=2$ this yields
\begin{align}
    \label{eq:polynomial_kernel}
    \left(\mathbf{K}^{(2)}\right)_{{ij}} &= k^{(2)}(\tilde{\mathbf{x}}_i, \tilde{\mathbf{x}}_j) \notag \\
    &= 1-\left(\tilde{\mathbf{x}}_i - \tilde{\mathbf{x}}_j\right)^2 + \frac{1}{2}\left(\tilde{\mathbf{x}}_i - \tilde{\mathbf{x}}_j\right)^4\,.
\end{align}
Our investigations encompass five different data encoding circuits from the literature~\cite{Hubregtsen.2022, Havlicek.2019,Haug_2023,Huang.2021,Slattery.2023} to define FQKs and PQKs. In our experiments, we chose the number of qubits of the data encoding circuits to match the respective feature dimension of the dataset, i.e., $n_{\text{q}} = n$. Moreover, we fix the number of layers to $L=2$ and use $\gamma = 1$ in Eq.~\eqref{eq:PQK_RBF} for the PQK definition to isolate the effect of bandwidth-tuning via $c$ and to avoid additional length-scale parameters. For details on the experimental setup, the functional forms of $k^{(d)}$ for $d>2$, and the encoding circuits we refer to Appendix~\ref{sec:appendix_circuits}.

In the following, we focus on findings corresponding to the \plasticcName\, dataset to clearly present the main results of our research. However, analogous simulations for the \kmnistName\, and \hiddenmanifoldName\, datasets, as demonstrated in Appendix~\ref{appendix:figures_for_bandwidth_tuned}, yield consistent and comparable results.

\begin{figure}[tb]
    \centering
    \includegraphics[width=0.54\linewidth]{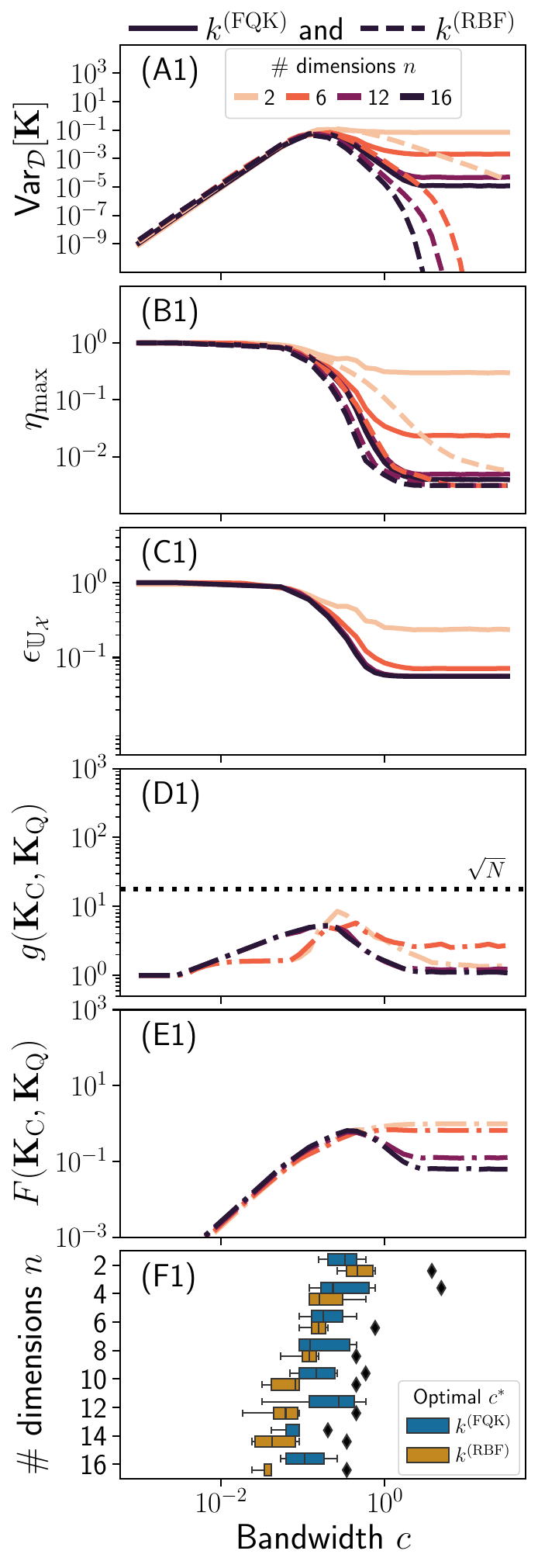}
    \includegraphics[width=0.4385\linewidth]{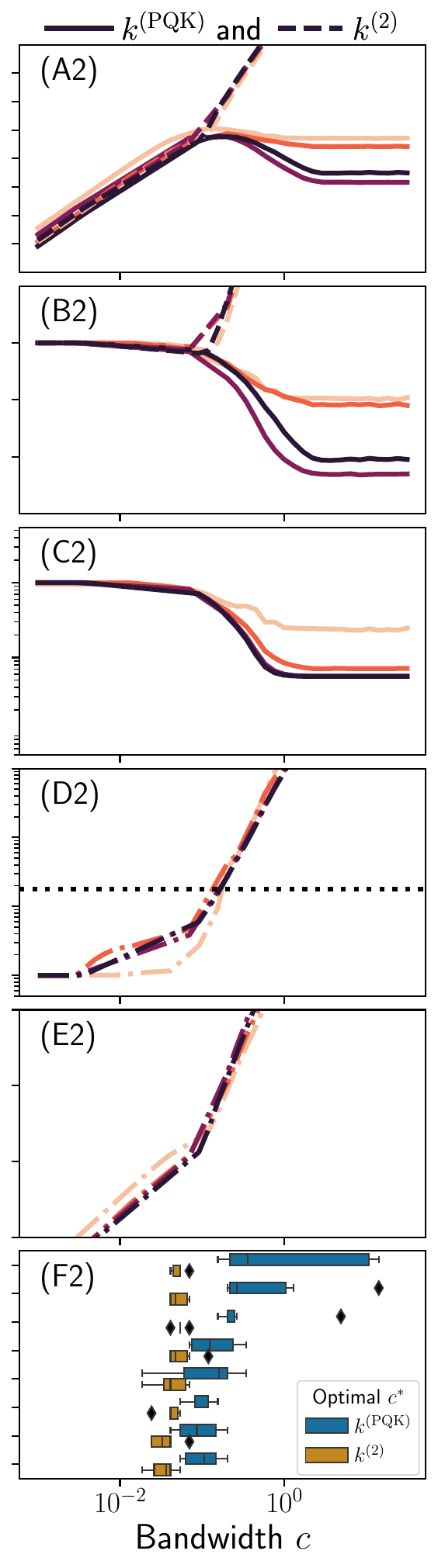}
    \caption{Effect of bandwidth-tuning on key quantities of QKs and CKs. The results are averaged over the different simulated seeds. The darker the color the larger the number of feature dimensions ($n=n_{\mathrm{q}}$). Solid lines represent QKs whereas dashed lines show to CKs. The left column corresponds to results with a FQK and $k^{\text{(RBF)}}$, while the right column shows PQK and $k^{\text{(2)}}$ results, both based on the encoding circuit from Ref.~\cite{Hubregtsen.2022}. (A1,2) show the variance of the kernel entries $\varKD$, (B1,2) the largest eigenvalue $\eta_{\mathrm{max}}$, and (C1,2) the expressivity of the underlying data encoding unitary $\epsilon_{\mathbb{U}_{\mathcal{X}}}$. (D1,2) shows the geometric difference $g(\mathbf{K}_C, \mathbf{K}_Q)$ according to Eq.~\eqref{eq:geo_diff_reg}, where the horizontal dashed line corresponds to $\sqrt{N}$, the limit where no quantum advantage is possible. (E1,2) gives the Frobenius distance $F(\mathbf{K}_C, \mathbf{K}_Q)$ according to Eq.~\eqref{eq:frobenius_distance}. 
    For clarity, we only assess the metrics $g$ and $F$ for FQKs with respect to $k^{\text{(RBF)}}$, and for PQKs with respect to $k^{\text{(2)}}$. 
    (F1,2) represents corresponding optimal bandwidths $c^*$ for different dimensions.}
    \label{fig:metrics_vs_bandwidth}
\end{figure}

\subsection{\label{sec:models_vs_c}Comparing key quantities of kernels as a function of bandwidth}

We address the question of which similarity measures bandwidth-tuned QKs define. To this end, we systematically compare distinguishing characteristics of CKs and QKs in terms of important properties and metrics to reveal their differences and 
commonalities. We compare FQKs and PQKs to RBF kernels. To unravel and understand the impact of the size of bandwidth tuning parameters on the similarity measure of QKs, we furthermore compare to polynomial kernels corresponding to truncated Taylor approximations of the RBF kernel.

We examine these quantities as a function of the bandwidth tuning parameter in Fig.~\ref{fig:metrics_vs_bandwidth}, where we show the average over all simulated seeds for QKs based on the encoding circuit from Ref.~\cite{Hubregtsen.2022} for different problem dimensions. The left column corresponds to results with a FQK and $k^{\text{(RBF)}}$, while the right column shows PQK and $k^{\text{(2)}}$ results. 
Besides the geometric difference $g$ of Eq.~\eqref{eq:geo_diff_reg}, to quantify the entry-by-entry difference between CKs $\mathbf{K}_{\mathrm{C}}$ and QKs $\mathbf{K}_{\mathrm{Q}}$, we moreover consider an asymmetric Frobenius distance measure
\begin{align}
\label{eq:frobenius_distance}
        F(\mathbf{K}_{\mathrm{C}},\mathbf{K}_{\mathrm{Q}}) &= \norm{\mathbf{K}_{\mathrm{C}} - \mathbf{K}_{\mathrm{Q}}}_F /  \norm{\mathbf{K}_{\mathrm{Q}}}_F \notag \\
        &=\frac{\sqrt{\sum_{ij} |k_{\mathrm{C}}(\tilde{\mathbf{x}}_i,\tilde{\mathbf{x}}_j) -  k_{\mathrm{Q}}(\tilde{\mathbf{x}}_i, \tilde{\mathbf{x}}_j)|^2}}{\sqrt{\sum_{ij} |k_{\mathrm{Q}}(\tilde{\mathbf{x}}_i, \tilde{\mathbf{x}_j})|^2}}\,.
\end{align}
For clarity, we only assess the metrics $g$ and $F$ for FQKs with respect to $k^{\text{(RBF)}}$, and for PQKs with respect to $k^{\text{(2)}}$. However, for $\varKD$, the largest eigenvalues $\eta_{\mathrm{max}}$ and the expressivity of the respective data encoding circuits $\epsilon_{\mathbb{U}_{\mathcal{X}}}$, the CK results can be swapped per column to allow comparing with the other QK.

First, we observe that each of the $\varKD$, $\eta_{\mathrm{max}}$, and $\epsilon_{\mathbb{U}_{\mathcal{X}}}$ results, i.e. Figs.~\ref{fig:metrics_vs_bandwidth}~(A1,2) to~(C1,2), yield three different regimes for both FQKs and PQKs that occur at small, intermediate, and large $c$ values.

For $\varKD$ in Figs.~\ref{fig:metrics_vs_bandwidth}~(A1,2) we obtain a polynomial decay for small bandwidths, which is mostly independent of the number of dimensions. Here, both RBF and polynomial kernels show a similar slope. In Sec.~\ref{sec:analytical_toy_model}, we show that the slope of the quantum models decays proportionally to $c^4$ in this regime. In the intermediate regime, the quantum models start to deviate from the polynomial models but still resemble the RBF kernel. The variance for QKs and RBF kernels peaks and then starts decreasing, while for the polynomial models it increases. This similarity vanishes for large bandwidths, where the QKs reach a plateau, while the polynomial models diverge. The plateaus of the QKs decrease with increasing dimensionality, except the ones for $n=12$ and $n=16$ of PQKs, which might be due to the limited dataset size. For the RBF, the variance decreases significantly.

For the largest eigenvalues in Figs.~\ref{fig:metrics_vs_bandwidth}~(B1,2), we observe $\eta_{\mathrm{max}}\to 1$ as $c\to 0$ for both quantum and classical models. For intermediate bandwidths, QKs and RBF kernels start to deviate from the polynomial models. Here, $\eta_{\mathrm{max}}$ of the QKs and RBF kernels decreases with increasing bandwidth, whereas the largest eigenvalues of polynomial kernels start to increase. In the large bandwidth limit, the polynomial models show completely different behavior from the RBF and QKs. The largest eigenvalues of polynomial kernels increase significantly, whereas the QKs show plateaus that decrease with increasing dimensionality. The RBF kernels progressively decrease up to a plateau value that asymptotically merges with the respective value of QKs for $(c,n)\to\infty$. For $n\to\infty$ (but finite $N$), the plateau of RBF and QKs converges to $\eta_{\mathrm{max}}=\nicefrac{1}{N}$, which corresponds to the largest eigenvalue of a $N \times N$ identity matrix.

The expressivity measure $\epsilon_{\mathbb{U}}$ in Figs.~\ref{fig:metrics_vs_bandwidth}~(C1,2) is exclusively defined for QKs as its formulation according to Eq.~\eqref{eq:expressiblity_definition} is based on data encoding circuits. Remarkably, expressivity and the largest eigenvalue exhibit a very similar functional form. Large bandwidth regimes correspond to most expressive encoding circuits, i.e. $\epsilon_{\mathbb{U}}\to 0$, while low bandwidths are associated with minimal expressivity. With increasing dimensionality, the plateau observed in the large bandwidth regime is related to the mean value of an $N \times N$ identity matrix, as similarly discussed for $\eta_{\text{max}}$.

The connection between the exponential concentration phenomenon, as measured by $\varKD$, the largest eigenvalue $\eta_{\mathrm{max}}$ of the kernel Gram matrix and the expressivity $\epsilon_{\mathbb{U}_{\mathcal{X}}}$ of corresponding encoding circuits can be explained by recent theoretical advancements. Ref.~\cite{Thanasilp.2024} proved that $\varKD$ is bounded by $\epsilon_{\mathbb{U}_{\mathcal{X}}}$. Similarly, Ref.~\cite{Kubler.} showed that the largest eigenvalue can be upper bounded by an expressivity-dependent term.

In regard to distance metrics, Fig.~\ref{fig:metrics_vs_bandwidth}~(D1), displays that independent of bandwidth the $\KFQK$ show small $g$ values when compared to $\kRBF$. Although a peak geometric difference exists for intermediate-sized bandwidths, the value is still below the one for potential quantum advantage ($g\propto\sqrt{N})$. For both small and large $c$ the geometric difference plateaus. This is in contrast to the geometric difference between $\KPQK$ and $\Kpoly{2}$ as depicted in Fig.~\ref{fig:metrics_vs_bandwidth}~(D2). Here, we can identify two distinct functional behaviors as a function of $c$. For small $c$-values, the geometric difference is also small, revealing the role of the bandwidth size to classical tractability. For large $c$-values, $g$ grows extremely fast, as the entries of $\Kpoly{2}$ increase as $c^4$; cf. Eq.~\eqref{eq:polynomial_kernel}. 
The results corresponding to the Frobenius distance according to Eq.~\eqref{eq:frobenius_distance} in Figs~\ref{fig:metrics_vs_bandwidth}~(E1,2), reveal that this metrics qualitatively shows the same functional behavior as observed for $g$ for both FQKs and PQKs. The observation that $F$ rapidly increases for $\KPQK$ and $\Kpoly{2}$ but not for the comparison between $\KRBF$ and $\KFQK$, is again due to the fact that the entries of $\Kpoly{2}$ increase as $c^4$ for large $c$. Thus in Fig.~\ref{fig:metrics_vs_bandwidth}~(D2) this does not point towards an overall quantum advantage of $\KPQK$ here. 

Within further comparison of other QKs with CKs in Appendix~\ref{appendix:heatmapsection}, we generally observe that QKs produce Gram matrices that closely resemble those of RBF kernels and in the small bandwidth regime, even those of the polynomial kernels.

Beyond that, the three bandwidth regimes serve as a valuable framework for comparing quantum models with $c$ set to the optimal value $c^*$. In this regard, the boxes showing $c^*$ in Figs.~\ref{fig:metrics_vs_bandwidth}~(F1,2) reveal that these always fall within the small- to intermediate-sized bandwidth regimes, where the properties are either very similar to RBF \emph{and} polynomial kernels or start to deviate from the polynomials models but are still close to the RBF results. The observation that QKs are always very similar to CKs in all the important properties and metrics analyzed here, suggests that the efficacy of bandwidth-tuned quantum models for the datasets examined can be eventually attributed to the classical behavior of the QKs in these regimes. In other words, bandwidth tuning on classical datasets avoids exponential concentration by restricting the Hilbert space to a smaller subspace, thus effectively decreasing the models expressivity (as measured by $\epsilon_{\mathbb{U}_{\mathcal{X}}}$) but enables generalization, as measured both by $\eta_{\mathrm{max}}$ and $\varKD$. In these regimes QKs effectively approximate CKs; either both RBF and polynomial kernels or only RBF kernels.

This finding is not restricted to the example shown in Fig~\ref{fig:metrics_vs_bandwidth} but, as shown in Appendix~\ref{appendix:quantities_vs_bandwidth}, can be consistently observed over other datasets as well. 
 
\subsection{\label{sec:analytical_toy_model}Analytical model}
To elucidate the underlying mechanism driving the three distinct regimes manifesting consistently across $\varKD (c,n)$, $\eta_{\mathrm{max}}(c,n)$, and $\epsilon_{\mathbb{U}_{\mathcal{X}}}(c,n)$ as observed in Fig.~\ref{fig:metrics_vs_bandwidth}, we derive a toy-model for FQKs that analytically captures the behavior of these properties. The detailed derivation of the expressions is given in Appendices~\ref{appendix:variance_kernel} and~\ref{appendix:expressivity}.

The analytical models are based on a FQK with a separable data encoding circuit made up of $L$ layers, each applying one-qubit rotations $R_X(x)$ on all $n_\mathrm{q}$ qubits (cf. Fig.~\ref{fig:rx_encoding}). This yields~\cite{Schuld.26012021, Kubler., canatatheorybandwidthenablesgeneralization}
\begin{equation}
    \label{eq:separable_R_X} 
    k_{R_X}(\tilde{\mathbf{x}}_i, \tilde{\mathbf{x}}_i^\prime) = \prod_{i=1}^{n_\mathrm{q}}\cos^2\left(\frac{L\left(\tilde{x}_{i}-\tilde{x}_{i}^\prime\right)}{2}\right)\,.
\end{equation}
We assume $N$ training points $\mathbf{x}_i\in\mathcal{X}$ each following an independent distribution $p(\mathbf{x}_i) = \prod_{j=1}^n p\left(x_{ij}\right)$ with each $x_{ij}$ distributed uniformly, i.e.,  $p(x_{ij}) =  1/(2\pi)$. When sampling training data  $x_{ij}\in \mathcal{X}=[-\pi, \pi]$, we obtain
\begin{equation}
\label{eq:toymodelmainuniform}
\begin{split}
    &\varksep (c,n) = \\
    & \left(\frac{3}{8} - \frac{\cos{\left(2 \pi Lc \right)}}{4 \pi^{2} (Lc)^{2}} - \frac{\cos{\left(4 \pi Lc \right)}}{64 \pi^{2} (Lc)^{2}} + \frac{17}{64 \pi^{2} (Lc)^{2}}\right)^{n} \\ & - \left({\frac{1}{2} - \frac{\cos{\left(2 \pi c \right)}}{4 \pi^{2} (Lc)^{2}} + \frac{1}{4 \pi^{2} (Lc)^{2}}}\right)^{2 n}\,.
\end{split}
\end{equation}
From this we quickly obtain the polynomial decaying behavior for $c\to 0$ as
\begin{equation}
\label{eq:small_c_limit_toy_model_uniform}
\varksep(c\rightarrow0,n) \approx  \alpha(n) (Lc)^4\,,
\end{equation}
with $\alpha(n) = \frac{7 \pi^4}{180}n$. For $c\to\infty$, this reduces to
\begin{align}
    \label{eq:large_c_limit_toy_model_uniform}
    \varksep(c\rightarrow \infty,n) &\approx \frac{a^{n} - b^{n}}{d^{n}}\,,
\end{align}
with $a = 3\cdot 2^4$, $\, b = 2^5, \, d = 2^7\,,$ which explains the appearance of plateaus in Figs.~\ref{fig:metrics_vs_bandwidth}~(A1,2) and therefore a behavior of $\varKD$ independent of $c$ in the large bandwidth regime. Furthermore, Eq.~\eqref{eq:large_c_limit_toy_model_uniform} captures the 
exponential decaying behavior of these plateaus with increasing dimensionality $n$\,.

\begin{figure}[tb]
    \centering
    \includegraphics[width=\linewidth]{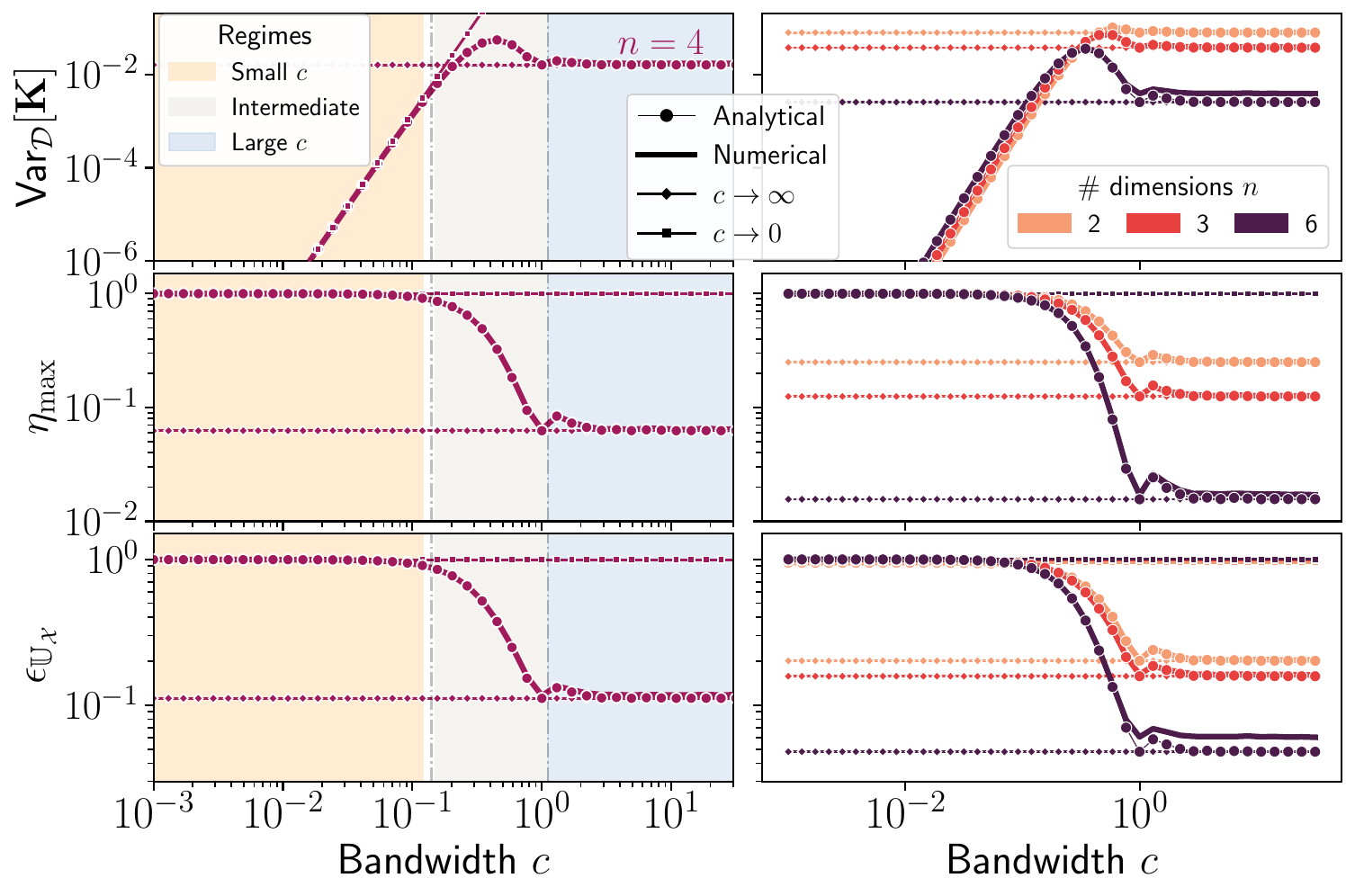}
    \caption{Overview of analytical model results corresponding to $\varKD$, $\eta_{\mathrm{max}}$, and $\epsilon_{\mathbb{U}_{\mathcal{X}}}$ as well as comparison to numerical results based on the $k_{R_x}$-kernel of Eq.~\eqref{eq:separable_R_X} with $L=2$. We assume a uniform distribution and sample $N=720$ data points in $x_{ij}\in \mathcal{X}=[-\pi, \pi]$ . From \textbf{top to bottom:} Variance of the kernel as a function of bandwidth, expressivity and largest eigenvalue. The \textbf{left} column corresponds to simulations with $n=4$ were the three distinct regimes are shaded for visual aid. The numerical results (blue solid lines) were obtained by sampling 720 points from the uniform distribution and calculating discrete quantities. For the analytical plots (orange dashed lines), we assume $N\rightarrow\infty$. The dotted lines represent the limiting behavior of the analytical quantities for the uniform distribution. The \textbf{right} column shows the quantities for different dimensionality.}
    \label{fig:analyticalmodelregimes}
\end{figure}

The largest eigenvalue $\eta_{\text{max}}(c, n)$ relies on the analytical expression for each eigenvalue of the kernel integral operator as derived in Ref.~\cite{canatatheoryclassicalkernels}. This defines $\eta_{\text{max}}(n, c)$ as given in Appendix~\ref{appendix:largesteigenvalue}.
with the corresponding limits given by, 
\begin{equation}
\label{eq:limit_expr}
\eta_{\text{max}} \overset{c \rightarrow 0}{=} 1\quad\text{and}\quad\eta_{\text{max}} \overset{c \rightarrow \infty}{=} \left( \nicefrac{1}{2} \right)^n\,.
\end{equation}

The expressivity as defined in Eq.~\eqref{eq:expressiblity_definition} can be formulated so that it depends only on the mean of the FQK Gram matrix squared and an additive constant dimensional term; cf. Appendix~\ref{appendix:expressivity}. From this one arrives at an analytical expression where the corresponding limits are given by
\begin{equation}
\label{eq:limit_expr}
\epsilon^2_{\mathbb{U}_{\mathcal{X}}} \overset{c \rightarrow 0}{=} 1+h\quad\text{and}\quad\epsilon^2_{\mathbb{U}_{\mathcal{X}}} \overset{c \rightarrow \infty}{=} \left( \nicefrac{3}{8} \right)^n+h\,,
\end{equation}
with the constant $h=\nicefrac{2 ! (2^n-1)!}{(2+2^n-1)!}$ from Eq.~\eqref{eq:expressivity_simplified_appendix}.

The quantities derived from the analytical model are shown in Fig.~\ref{fig:analyticalmodelregimes} for $N=720$ data points sampled uniformly in $x_{ij}\in \mathcal{X}=[-\pi, \pi]$ together with their corresponding limits for $c\to 0$ and $c\to\infty$, respectively. We support their validity by comparing to numerical results obtained from using the data distribution in Eq.~\eqref{eq:separable_R_X}. The numerical variance was obtained using Eq.~\eqref{eq:variance_discrete} (including the diagonal). For the expressivity, we average the values of the obtained $\KFQK$ squared and add the constant dimensionality term; cf. Appendix~\ref{appendix:expressivity}. The largest eigenvalues are obtained from diagonalizing $\KFQK$. Both numerical experiments and analytical calculations align given enough sample points. The analytical limits accurately capture the limiting behavior observed in Fig.~\ref{fig:metrics_vs_bandwidth}. In the right column of  Fig.~\ref{fig:analyticalmodelregimes} we further demonstrate that the analytical model is also able to qualitatively capture the dependence on the dimensionality $n$. For higher dimensions (in this case $n\geq 6$), the discrete nature of the sampling size affects the limiting behavior but preserves the overall functional form. In this higher-dimensional regime, the number of encoded unitaries is insufficient to saturate the Hilbert space; thus, the analytical limits are not reached. This is shown in Appendix~\ref{appendix:effect_of_finite_number}, where we thoroughly analyze this dependence on the number of data points. Furthermore, increasing the number of layers leads to a general shift of the regimes to the left, such that the maximum value of the variance occurs for lower bandwidths. In the simple analytical models this can be easily tracked, as the number of layers $L$ simply acts as an additional rescaling factor in Eq.~\eqref{eq:separable_R_X}. This is detailed in Appendix~\ref{appendix:effect_of_layers}, where we demonstrate the occurrence of the shift for other datasets and encoding circuits.

Comparing Figs.~\ref{fig:metrics_vs_bandwidth} and~\ref{fig:analyticalmodelregimes}, we note that despite its simplicity and derivation based on FQKs with separable encoding circuits, the analytical models also qualitatively capture the behavior of these properties for FQKs and PQKs across more complex encoding circuits as well as for non-trivial data distributions.

\subsection{\label{sec:models_vs_n}Comparison of optimally bandwidth-tuned quantum and classical kernels}

\begin{figure*}[tb]
    \centering
    \includegraphics[width=1\linewidth]{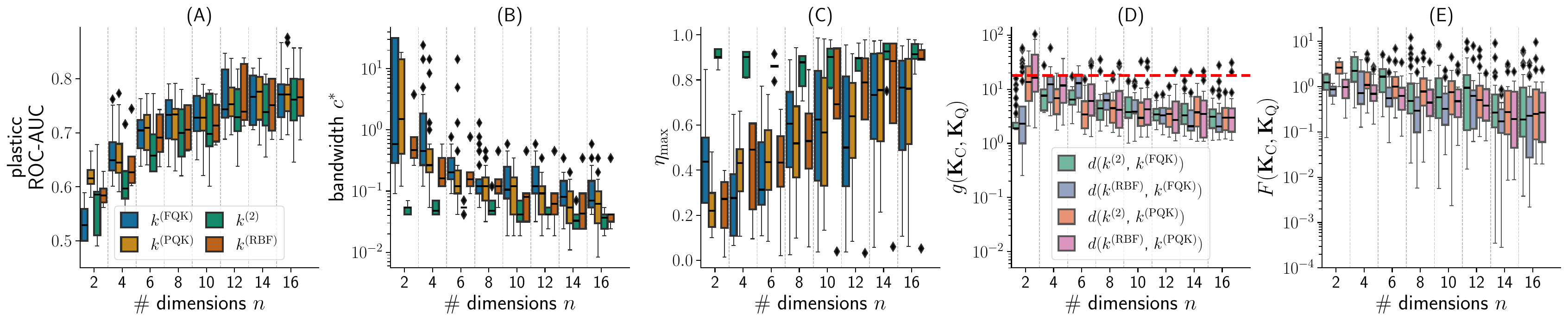}
    \caption{Comparison of QKs and CKs for important properties and metrics as a function of increasing feature dimensionality ($n=n_{\mathrm{q}}$). As QKs we consider $\kFQK$ (blue) and $\kPQK$ (yellow), and as CKs we consider $\kpoly{2}$ (green) and $\kRBF$ (brown). Panels (A) -- (C) depict ROC-AUC score, optimal bandwidth $c^*$, and largest eigenvalue $\eta_{\mathrm{max}}$, respectively. For panels (D) and (E), we report the geometric difference $g(K_C, K_Q)$ and Frobenius distance $F(K_C, K_Q)$, respectively. Here we use $d(\cdot,\cdot)$ in the legend to label both $g(\cdot,\cdot)$ and $F(\cdot,\cdot)$, where green and orange (blue and pink) boxes refer to comparisons between the QKs with $\kpoly{2}$ ($\kRBF$). The results are aggregated across optimal bandwidths $c^*$ and regularization, yielding the highest ROC-AUC scores for each seed. The QK results are moreover aggregated over all data encoding circuits. The red dashed line for the geometric difference corresponds to $\sqrt{N}$.}
    \label{fig:quantities_vs_n}
\end{figure*}

Equipped with the findings of the previous sections, we take a closer look at the characteristics of optimally bandwidth-tuned quantum and classical kernels and detail the role of increasing feature dimension $n$. The corresponding simulation results are shown in Fig.~\ref{fig:quantities_vs_n}. Here, we report aggregated findings across the different data encoding circuits with corresponding optimal bandwidths $c^*$ and optimal regularization parameters yielding the highest receiver operating characteristic curve (ROC-AUC) scores on each of the samples. We compare these findings to optimally bandwidth-tuned $\kpoly{2}$ and $\kRBF$ kernels. 
The analogous study for the \kmnistName\, and \hiddenmanifoldName \, \,datasets are shown in Appendix~\ref{appendix:quantities_vs_n}. We generally observe the same behavior as discussed below.

As shown in Fig.~\ref{fig:quantities_vs_n}~(A), as the number of dimensions increases, the ROC-AUC scores on the test dataset of both classical and quantum methods equally improve, with similar predictions scores between both methods for each $n$.
Furthermore, Fig.~\ref{fig:quantities_vs_n}~(B), similar to Figs.~\ref{fig:metrics_vs_bandwidth}~(F1,2), shows that in the case of the \plasticcName\, dataset, higher dimensions require smaller optimal bandwidth parameters $c^*$ to obtain optimal model performance. Again, the respective behavior of CKs and QKs is very similar. Except for the polynomial models, which already have small $c^*$ for low dimensions. 

The size of the bandwidths directly affects the spectra of the corresponding Gram matrices. The largest eigenvalues $\eta_{\mathrm{max}}$ of the Gram matrices (cf. Sec~\ref{sec:background}) are shown in Fig.~\ref{fig:quantities_vs_n}~(C). Their non-decaying behavior explains the propensity for generalization in Fig.~\ref{fig:quantities_vs_n}~(A). Moreover, with decreasing optimal bandwidths and thus increasing $n$, we observe that $\eta_{\mathrm{max}}\approx 1$ for CKs and QKs. For $\kpoly{2}$, this can already be seen for low dimensions. Remarkably,  $\eta_{\mathrm{max}}$ of the RBF kernels aligns with corresponding $\eta_{\mathrm{max}}$ of QKs across all orders of magnitude and dimensions. This also holds for the full spectral shape as shown in Appendix~\ref{appendix:eigenvalues}. As $n$ increases and the $c^*$ values of the kernels decrease, we observe similar spectra for $\kpoly{2}$, $\kRBF$, FQKs and PQKs.

The behavior and size of the bandwidths are also reflected in the geometric difference between the optimal CKs and QKs as shown in Fig.~\ref{fig:quantities_vs_n}~(D). As $c$ decreases for higher dimensions, $g(\mathbf{K}_C,\mathbf{K}_Q)$ decreases as well. Since the median $g$ values of all models are below the size required for quantum advantage (i.e., $g\propto\sqrt{N}$) we are guaranteed that the classical model provides similar or better performance~\cite{Huang.2021}. Again, this explains Fig.~\ref{fig:quantities_vs_n}~(A). Interestingly, Fig.~\ref{fig:quantities_vs_n}~(D) moreover uncovers, that both bandwidth-tuned PQKs and FQKs are classically replaceable not only by $\kRBF$ but also by its simple second-order Taylor expansion $\kpoly{2}$. 

These connections between optimal performance scores, bandwidth-tuning parameters and geometric differences align with observations reported in Refs.~\cite{Slattery.2023, Egginger.2024}, where it was numerically shown that accuracy and geometric difference depend in an opposing manner on bandwidth tuning. This can be tracked by investigating their Frobenius distance, cf. Eq.~\eqref{eq:frobenius_distance}. As shown in Fig.~\ref{fig:quantities_vs_n}~(E) both quantum and classical kernel matrices are not only similar with regards to their generalization properties, but also the distance between kernel Gram matrix entries reflects the previous observations. As the optimal bandwidth decreases, the Frobenius distance decreases as well. We observe the same behavior for the remaining datasets of this study as shown in Appendix~\ref{appendix:heatmapsection}.

\section{\label{sec:discussion}Discussion}
The findings of our study reveal why and how bandwidth-tuned QKs become classically tractable by thoroughly comparing to conventional SVMs based on RBF and polynomial kernels. The bandwidth tuning parameters yielding optimal model performance for the datasets considered in this work, are so small in size that QKs and CKs become similar. We provide empirical evidence for this by calculating the geometric difference and a Frobenius-norm-based distance measure as well as for the ROC-AUC scores, kernel spectra and variance. Our investigations include QK models based on both FQKs and PQKs with different data sets and various encoding circuits from common QML literature. This significantly extends previous works~\cite{Slattery.2023, canatatheorybandwidthenablesgeneralization, Shaydulin.2022} and allows for more general insights.

In particular, we contribute to understand the similarity measure induced by QKs in high dimensions as raised in Ref.~\cite{Bowles.11032024}. Besides the observation that RBF kernels and QKs show comparable performance and have small Frobenius distance, we additionally find that they exhibit comparatively small geometric differences across a large range of bandwidth sizes. For optimal bandwidths, we reveal that this resemblance is even more pronounced and is also shared by the kernel spectra and variance. Since the size of the optimal bandwidths is so small, we numerically demonstrate that measures of similarity defined by both FQKs and PQKs simplify even further and effectively behave like polynomial kernels corresponding to low-order Taylor approximations of RBF kernels~\cite{inria-00325810, RING2016107, cotter2011explicitapproximationsgaussiankernel}.

Furthermore, our simulations unveil that the variance $\varKD$, the largest eigenvalue of QKs $\eta_{\mathrm{max}}$ as well as the corresponding expressivity $\epsilon_{\mathbb{U}_{\mathcal{X}}}$ of the underlying encoding circuit generally share three distinct regimes for small, intermediate and large values of the bandwidth tuning. For the datasets of this study, optimal $c$-values always belong to the small to intermediate regime. This demonstrates how exponential concentration effects are avoided, but also how QKs are effectively dequantized. The behavior of these quantities can be analytically modeled using a FQK based on a separable data encoding. Despite its simplicity, it captures the behavior of these properties for both FQKs and PQKs across various encoding circuits as well as for all datasets studied in this work. The model correctly predicts an exponential decay of $\varKD$ in the large $c$-regime as the number of qubits increases; as reported in Refs.~\cite{Thanasilp.2024, Kubler., Huang.2021}. This dependence on the problem dimensionality vanishes in the small-$c$-regime, where $\varKD$ decays polynomially in $c$. 

In the small to intermediate regimes, where the optimal quantum models belong to, we observe reasonably large values for $\varKD$ and non-decaying largest eigenvalues $\eta_{\mathrm{max}}$, which explains good generalization behavior~\cite{canatatheorybandwidthenablesgeneralization, Kubler., canatatheoryclassicalkernels}. However, the classical models behave similar in these regimes. Additionally, our experiments reveal that the intermediate and small-bandwidth regime correspond to regions where the geometric differences are small. Thus, our findings clearly establish a connection between exponential concentration (as measured by $\varKD$), generalization properties (as measured by the largest eigenvalue) and classical tractability (as measured through $g$) within the framework of bandwidth tuning.

Beyond that, our study contributes to an interesting analogy between recent theoretical advancements in QML, indicating that the absence of barren plateaus implies classical simulability~\cite{CerezoAbsence}, e.g., in quantum convolutional neural networks~\cite{Pesah.2021, Bermejo.}. Exponential concentration constitutes a similar barrier for QK methods~\cite{Thanasilp.2024, gilfuster2025relationtrainabilitydequantizationvariational}. While bandwidth tuning alleviates this barrier and increases the model performance, our work emphasizes that this is due to the fact that both FQK and PQK models align more closely with those classical ones, effectively resulting in a dequantization.

We note that albeit we use different datasets and consider different model designs, the sample size still represents a restriction, which limits the generalizability of our results. For instance, the analytical model could reproduce the $c$-dependence in all QK models, except for PQKs based on the encoding circuit from Ref.~\cite{Havlicek.2019}. Future studies should carefully investigate this. Furthermore, it should extend this to larger sample sizes and analyze the dependence on dataset complexity. While our main conclusion that RBF kernels or even polynomial kernels can effectively capture the similarity measure induced by bandwidth-tuned QKs applies to all datasets investigated in this study, this finding is generally dataset-dependent as indicated in Fig.~\ref{fig:heatplotmapsdifferences} of Appendix~\ref{appendix:heatmapsection}.

Apart from that, we believe that another important avenue for future research should center on identifying proper datasets and mechanisms that fall outside the scheme discussed here and bear the potential that may lead to quantum advantage. While we observed a clear dataset dependence in our study, bandwidth-tuning corresponds to a rather trivial way of artificially modifying data and it is not clear which dataset properties should be present to fully leverage the quantum-specific advantages. A brute-force approach could be via artificial datasets, by maximizing the geometric (and/or other) difference(s) or by exploiting the Fourier properties of quantum models, as for example indicated in Refs.~\cite{Sweke_2025_potentialandlimitation, sweke2025kernelbaseddequantizationvariationalqml} or study potential advantages through the ETK lens~\cite{shin2025newperspectivesquantumkernels}. The second layer to this is to systematically investigate how to construct data-specific encoding circuits. Here, algorithmically-driven circuit generation, cf., e.g., Refs.~\cite{Altares-López_2021, dai2024quantummachinelearningarchitecture, Rapp_2025} could be an interesting research direction. Overall, we believe that our findings corroborate the need to tackle QML research from two sides, i.e. identify datasets that require to leverage quantum-specific properties and explore how corresponding model design should look like.

\vspace*{2em}
\section{\label{sec:conclusion} Conclusion}
This work demonstrates how both FQKs and PQKs can become classically tractable using common data encoding circuits and datasets, showing that optimally bandwidth-tuned QKs essentially behave like classical RBF kernels. This further simplifies as the size of these bandwidths can be so small that QKs effectively behave like polynomial kernels corresponding to low-order Taylor approximations of RBF kernels. Although this prevents exponential concentration and therefore enables generalization, it renders QKs classically tractable. We found that key performance indicators like kernel spectrum align between QKs and CKs. Additionally, we identified three distinct variance regimes in QKs as a function of bandwidth that also emerge in the largest eigenvalue and data encoding expressivity. Our analytical model, based on a FQK with a separable encoding circuit, captures the behavior of these quantities for both FQKs and PQKs, also with more complex data encoding circuits across different bandwidths and problem dimensions.

\section*{Acknowledgments}
This work was supported by the German Federal Ministry of Education and Research through the project H2Giga Degrad-EL\textsuperscript{3} (grant no. 03HY110D). The authors acknowledge Jörg Main and the Institute for Theoretical Physics I of the University of Stuttgart for supervising R.F.A Master's thesis, from which this paper was partly derived.

\section*{Data availability statement}
The code to reproduce all of the simulation results of this study is available via GitHub~\cite{robertogit}. Data used in this study is publicly available and can be accessed via the respective references provided in the manuscript.

%\bibliography{bibliography}

\clearpage

\appendix
% Reset page counter and redefine pager numbering
\setcounter{page}{1}
\renewcommand{\thepage}{A\arabic{page}}
\setcounter{figure}{0}
\renewcommand{\thefigure}{A\arabic{figure}}

\section{Experimental details}
\label{sec:appendix_experimental_details}

For both QKs and CKs, we performed a grid search over the following regularization parameters $C=[32.0, 64.0, 128.0, 512.0, 1024.0]$ of the SVMs, and used 40 uniformly separated bandwidths in the log scale over the interval $c \in [10^{-3}, 10^{1.5}]$. For the geometric difference, we used $\lambda = \nicefrac{1}{2C}$ and for the Gram matrix diagonalization, we normalized by the number of data points, i.e. $\text{eig}(\nicefrac{\mathbf{K}}{N})$. The datasets are preprocessed as described in the main text, i.e. train-test split, standardization (remove the mean) and PCA except for the \plasticcName \, dataset from Ref.~\cite{team.28092018}, which is already standardized, wherefore we skip this step.

The order $2t$ polynomial kernels are obtained from a Taylor expansion of the RBF kernel according to Eq. \ref{eq:RBF} (with $\gamma=1$) truncated after the $t$-th term
\begin{equation}\label{eq:appendix_poly1}
    \left(\mathbf{K}^{(t)}\right)_{{ij}} = k^{(t)}(\tilde{\mathbf{x}}_i, \tilde{\mathbf{x}}_j) = \sum_{t=0}^d (-1)^t\frac{\left(\tilde{\mathbf{x}}_i - \tilde{\mathbf{x}}_j \right)^{2t}}{t!}\,.
\end{equation}

\subsection{Data encoding circuits}
\label{sec:appendix_circuits}

Here, we depict the data encoding circuits used in this study. All the circuits are chosen from common QML literature and are available in sQUlearn \cite{kreplin2025squlearn}. For those circuits with additional (trainable) parameters $\mathbf{p}$ besides the encoding variables $x$, we randomly initialized $\mathbf{p}$ with seed 1. 

\subsection{Separable $R_x$}
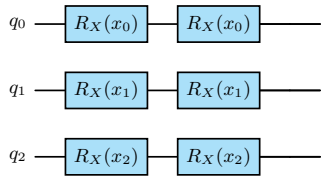
\begin{figure}[h]
    \centering
    \scalebox{0.80}{
     \begin{quantikz}
         \lstick{$q_0$} & \gate[style={fill=cyan!30}]{R_X(x_0)} & \gate[style={fill=cyan!30}]{R_X(x_0)} & \qw & \qw \\
         \lstick{$q_1$} & \gate[style={fill=cyan!30}]{R_X(x_1)} & \gate[style={fill=cyan!30}]{R_X(x_1)} & \qw & \qw \\
         \lstick{$q_2$} & \gate[style={fill=cyan!30}]{R_X(x_2)} & \gate[style={fill=cyan!30}]{R_X(x_2)} & \qw & \qw \\
     \end{quantikz}
     }
     \caption{Data encoding circuit for the Separable $R_X$ from Refs. \cite{Kubler.,Schuld.26012021,canatatheorybandwidthenablesgeneralization} with $n=3$ and $L=2$.}
     \label{fig:rx_encoding}
\end{figure}

This encoding circuit has been largely explored in the literature~\cite{Kubler.,Schuld.26012021,canatatheorybandwidthenablesgeneralization} given that the kernel produced by it can be analytically written as, $k^{\text{FQK}}(\tilde{\mathbf{x}}, \tilde{\mathbf{x}}^\prime) = \prod_{i=1}^{n_\text{q}} \cos^2\left(\frac{L\left(\tilde{\mathbf{x}}_i-\tilde{\mathbf{x}}_i^\prime\right)}{2}\right)$ for FQKs and $k^{\text{PQK}}(\tilde{\mathbf{x}}, \tilde{\mathbf{x}}^\prime, c) = e^{ -2 \gamma \left(  n_\text{q} - \sum_k^{n_q} \cos{(L(\tilde{\mathbf{x}}_k-\tilde{\mathbf{x}}_k^\prime)) } \right)}$ for RBF-PQKs. An example with two layers and three qubits is shown in Fig.~\ref{fig:rx_encoding}.

\subsection{Instantaneous quantum polynomial (IQP) embedding}
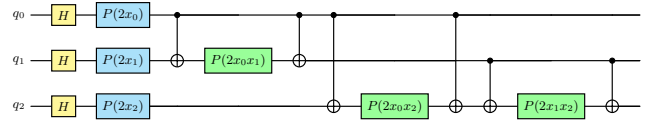
\begin{figure}[h]
    \centering
    \scalebox{0.55}{ 
    \setlength{\tabcolsep}{0.8em} 
    \begin{quantikz}
        \lstick{$q_0$} & \gate[style={fill=yellow!50}]{H} & \gate[style={fill=cyan!30}]{P(2x_0)} & \ctrl{1} & \qw & \ctrl{1} & \ctrl{2} & \qw & \ctrl{2} & \qw & \qw & \qw & \qw \\
        \lstick{$q_1$} & \gate[style={fill=yellow!50}]{H} & \gate[style={fill=cyan!30}]{P(2x_1)} & \targ{} & \gate[style={fill=green!40}]{P(2x_0x_1)} & \targ{} & \qw & \qw & \qw & \ctrl{1} & \qw & \ctrl{1} & \qw \\
        \lstick{$q_2$} & \gate[style={fill=yellow!50}]{H} & \gate[style={fill=cyan!30}]{P(2x_2)} & \qw & \qw & \qw & \targ{} & \gate[style={fill=green!40}]{P(2x_0x_2)} & \targ{} & \targ{} & \gate[style={fill=green!40}]{P(2x_1x_2)} & \targ{} & \qw 
    \end{quantikz}
    }
    \caption{Instantaneous Quantum Polynomial (IQP) data encoding circuit from Ref.~\cite{Harrow.2009} with $n=3$ and $L=1$.}
    \label{fig:IQP_ZZ_embedding}
\end{figure}

The IQP circuit is inspired by the encoding circuits of Ref.~\cite{Havlicek.2019} and was also used in the Refs.~\cite{Shaydulin.2022, canatatheorybandwidthenablesgeneralization} within the context of bandwidth tuning. The IQP circuit, implements the following mapping
\begin{equation}\label{eq:IQPembedding}
    |\phi(\tilde{\mathbf{x}})\rangle=\hat{U}_z(\tilde{\mathbf{x}})\hat{H}^{\otimes d}\hat{U}_z(\tilde{\mathbf{x}})\hat{H}^{\otimes d}|0\rangle\,,
\end{equation}
with 
\begin{equation}
     \hat{U}_{Z}(\tilde{\mathbf{x}})=\exp{\sum_{j=1}^{d} \tilde{\mathbf{x}}_j\hat{Z}_{j}+\sum_{j,k=1}^{d}\tilde{\mathbf{x}}_j \tilde{\mathbf{x}}_k \hat{Z}_{j}\hat{Z}_{k}}\,.
\end{equation} 
The output of this distribution is believed to be hard to simulate classically~\cite{Havlicek.2019, goldberg2017complexityapproximatingcomplexvaluedising}. In Fig.~\ref{fig:IQP_ZZ_embedding}, we show an example with one layer and and three qubits.

\subsection{HZY\_CZ circuit}
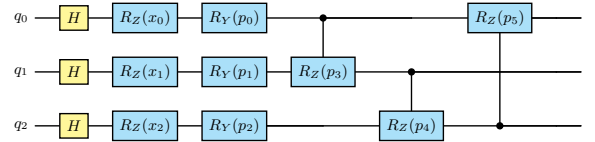
\begin{figure}[h]
    \centering
    \scalebox{0.65}{
    \begin{quantikz}
        \lstick{$q_0$} & \gate[style={fill=yellow!50}]{H} & \gate[style={fill=cyan!30}]{R_Z(x_0)} & \gate[style={fill=cyan!30}]{R_Y(p_0)} & \ctrl{1} & \qw & \gate[style={fill=cyan!30}]{R_Z(p_5)} & \qw & \qw \\
        \lstick{$q_1$} & \gate[style={fill=yellow!50}]{H} & \gate[style={fill=cyan!30}]{R_Z(x_1)} & \gate[style={fill=cyan!30}]{R_Y(p_1)} & \gate[style={fill=cyan!30}]{R_Z(p_3)} & \ctrl{1} & \qw & \qw & \qw \\
        \lstick{$q_2$} & \gate[style={fill=yellow!50}]{H} & \gate[style={fill=cyan!30}]{R_Z(x_2)} & \gate[style={fill=cyan!30}]{R_Y(p_2)} & \qw & \gate[style={fill=cyan!30}]{R_Z(p_4)} & \ctrl{-2} & \qw & \qw \\
    \end{quantikz}
    }
    \caption{Data encoding circuit HZY\_CZ from Ref.~\cite{Hubregtsen.2022} with $n=3$ and $L=1$.}
    \label{fig:hubregtsenencodingcircuitHZY}
\end{figure}

The HZY\_CZ data encoding circuit, introduced in Ref.~\cite{Hubregtsen.2022}, begins with a Hadamard layer, followed by an encoding layer of $R_Z$ rotations. This is succeeded by a series of parametrized rotations, which include both entangled and non-entangled operations. While the original implementation in Ref.~\cite{Hubregtsen.2022} optimized the parameters using kernel-target alignment, in this work, we randomly select the parameter values $\mathbf{p}$ and keep them fixed. An example with one layer and three qubits is shown in Fig.~\ref{fig:hubregtsenencodingcircuitHZY}.

\subsection{YZ\_CX circuit}
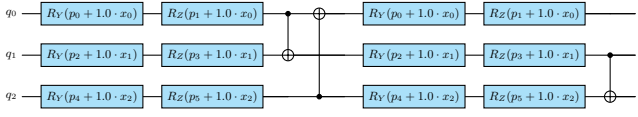
\begin{figure}[h]
    \centering
    \scalebox{0.50}{
    \begin{quantikz}
        \lstick{$q_0$} & \gate[style={fill=cyan!30}]{R_Y(p_0 + 1.0 \cdot x_0)} & \gate[style={fill=cyan!30}]{R_Z(p_1 + 1.0 \cdot x_0)} & \ctrl{1} & \targ{}   & \qw & \gate[style={fill=cyan!30}]{R_Y(p_0 + 1.0 \cdot x_0)} & \gate[style={fill=cyan!30}]{R_Z(p_1 + 1.0 \cdot x_0)}  & \qw & \qw  \\
        \lstick{$q_1$} & \gate[style={fill=cyan!30}]{R_Y(p_2 + 1.0 \cdot x_1)} & \gate[style={fill=cyan!30}]{R_Z(p_3 + 1.0 \cdot x_1)} & \targ{}  & \qw       & \qw & \gate[style={fill=cyan!30}]{R_Y(p_2 + 1.0 \cdot x_1)} & \gate[style={fill=cyan!30}]{R_Z(p_3 + 1.0 \cdot x_1)} & \ctrl{1} & \qw  \\
        \lstick{$q_2$} & \gate[style={fill=cyan!30}]{R_Y(p_4 + 1.0 \cdot x_2)} & \gate[style={fill=cyan!30}]{R_Z(p_5 + 1.0 \cdot x_2)} & \qw      & \ctrl{-2} & \qw & \gate[style={fill=cyan!30}]{R_Y(p_4 + 1.0 \cdot x_2)} & \gate[style={fill=cyan!30}]{R_Z(p_5 + 1.0 \cdot x_2)} & \targ{} & \qw  \\
    \end{quantikz}
    }
    \caption{Data encoding circuit YZ\_CX from Ref.~\cite{Haug_2023} with $n=3$ and $L=2$. }
    \label{fig:yzcsencodingcircuit}
\end{figure}

Inspired by the encoding circuit used in Ref. \cite{Haug_2023} to perform learning tasks on large datasets using randomized measurements, we use the implemented sQUlearn \cite{kreplin2025squlearn} version that removes the randomized sampling of rotations at the circuit end. In Fig.~\ref{fig:yzcsencodingcircuit}, we show an example with two layers and three qubits.

\subsection{Z embedding circuit}
We use the sQUlearn implementation of the Qiskit~\cite{javadiabhari2024quantumcomputingqiskit} \texttt{ZFeatureMap}. It consists of Hadamard layers followed by encoded rotation around the $z$-axis and entangling gates. 
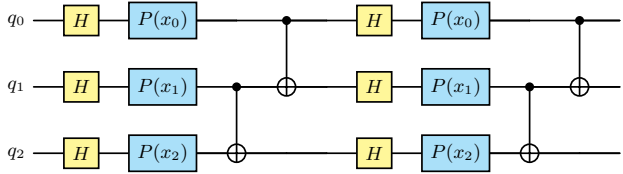
\begin{figure}[h]
    \centering
    \scalebox{0.80}{
    \begin{quantikz}
\lstick{$q_0$} & \gate[style={fill=yellow!50}]{H} & \gate[style={fill=cyan!30}]{P(x_0)} & \qw & \ctrl{1} & \qw & \gate[style={fill=yellow!50}]{H} & \gate[style={fill=cyan!30}]{P(x_0)} & \qw & \ctrl{1} & \qw   \\
\lstick{$q_1$} & \gate[style={fill=yellow!50}]{H} & \gate[style={fill=cyan!30}]{P(x_1)} & \ctrl{1} & \targ & \qw  & \qw & \gate[style={fill=yellow!50}]{H} & \gate[style={fill=cyan!30}]{P(x_1)} & \ctrl{1} & \targ & \qw  & \qw  \\
\lstick{$q_2$} & \gate[style={fill=yellow!50}]{H} & \gate[style={fill=cyan!30}]{P(x_2)} & \targ & \qw  & \qw & \qw & \gate[style={fill=yellow!50}]{H} & \gate[style={fill=cyan!30}]{P(x_2)} & \targ & \qw  & \qw & \qw   \\
\end{quantikz}
    }
    \caption{Data encoding circuit Z embedding with $n=3$ and $L=2$ as implemented in sQUlearn~\cite{kreplin2025squlearn}.}
    \label{fig:Zembedding}
\end{figure}

\section{Analytic derivation of variance, expressivity and largest eigenvalue \label{appendix:analytical_derivations} } 
Here, we provide details on the analytical derivations reported on Sec.~\ref{sec:analytical_toy_model}.

 \subsection{Variance}
The variance of a random variable $k(\mathbf{x},\mathbf{x}^\prime)$ over the variation of $\mathbf{x}$ and $\mathbf{x}^\prime$ sampled from $p(\mathbf{x})$ and $p(\mathbf{x}^\prime)$ is described by,
\begin{equation}
     \label{appendix:variance_kernel}
     \text{Var}_{\mathbf{x}, \mathbf{x}^\prime}[k(\mathbf{x},\mathbf{x}^\prime)]= \mathbb{E}_{\mathbf{x}, \mathbf{x}^\prime} \left[ k(\mathbf{x}, \mathbf{x}^\prime)^2 \right] - \mathbb{E}_{\mathbf{x}, \mathbf{x}^\prime} \left[ k(\mathbf{x},\mathbf{x}^\prime) \right]^2\,,
\end{equation}
where the expectation value is given by,
\begin{equation}
    \label{appendix:mean_K}
    \mathbb{E}_{\mathbf{x}, \mathbf{x}^\prime} \left[ k(\mathbf{x},\mathbf{x}^\prime) \right] = \int_{\mathbf{x}, \mathbf{x}^\prime} k(\mathbf{x},\mathbf{x}^\prime) p(\mathbf{x}) p(\mathbf{x}^\prime) \mathrm{d}\mathbf{x} \mathrm{d}\mathbf{x}^\prime\,.
\end{equation}
Plugging $k_{R_X}^{\text{FQK}}$ (Eq.~\eqref{eq:separable_R_X}) into Eqs.~\eqref{appendix:variance_kernel} and~\eqref{appendix:mean_K}, then performing exact integration leads to 
\begin{equation}
\label{eq:toymodelmainuniformappendix}
\begin{split}
    &\varksep (c,n) = \\
    & \left(\frac{3}{8} - \frac{\cos{\left(2 \pi Lc \right)}}{4 \pi^{2} (Lc)^{2}} - \frac{\cos{\left(4 \pi Lc \right)}}{64 \pi^{2} (Lc)^{2}} + \frac{17}{64 \pi^{2} (Lc)^{2}}\right)^{n} \\ & - \left({\frac{1}{2} - \frac{\cos{\left(2 \pi L c \right)}}{4 \pi^{2} (Lc)^{2}} + \frac{1}{4 \pi^{2} (Lc)^{2}}}\right)^{2 n}\,.
\end{split}
\end{equation}

For $c \rightarrow 0$, we perform a Taylor expansion of Eq.~\eqref{eq:toymodelmainuniformappendix} in $c$ around zero, up to fourth order, yielding Eq.~\eqref{eq:small_c_limit_toy_model_uniform} in the main text. On the other hand, for $c\rightarrow \infty$, all terms with $\frac{1}{(Lc)^2}$ in Eq.~\eqref{eq:toymodelmainuniformappendix} vanish as the cosine functions are bounded, obtaining, 
\begin{equation}
    \varksep (c,n) = \left(\frac{3}{8} \right)^2 - \left(\frac{1}{2} \right)^{2n},
\end{equation}
which can be easily rewritten as Eq.~\eqref{eq:large_c_limit_toy_model_uniform} of the main text.

We note that the exact integration of Eq.~\eqref{appendix:variance_kernel} for $n$ qubits is possible for $k_{R_X}^{\text{FQK}}$, as $k_{R_X}^{\text{FQK}}(\mathbf{x}_i, \mathbf{x}_i^\prime) = \prod^n_i k(x_i, x_i^\prime)$, with $k(x_i, x_i^\prime)$ the 1-qubit kernel. Then, assuming independent distributions for each coordinate, $p(\mathbf{x}) = \prod_{i=1} p(x_i)$, we find,
\begin{align}
    &\varksep = \notag \\ &=\mathbb{E}_{\mathcal{\mathbf{x}, \mathbf{x}^\prime}} \left[ k_{R_X}(\mathbf{\tilde{x}}, \mathbf{\tilde{x}}^\prime)^2 \right] - \mathbb{E}_{\mathcal{\mathbf{x}, \mathbf{x}^\prime}} \left[ k_{R_X}(\mathbf{\tilde{x}}, \mathbf{\tilde{x}}^\prime) \right]^2 \\ &= \prod^n_i \mathbb{E}_{\mathcal{\mathbf{x}, \mathbf{x}^\prime}} \left[  k(\tilde{x}_i, \tilde{x}_i^\prime)^2 \right] -  \prod^n_i \left( \mathbb{E}_{\mathcal{\mathbf{x}, \mathbf{x}^\prime}} \left[  k(\tilde{x}_i, \tilde{x}_i^\prime) \right] \right)^2\,.
\end{align}

\subsection{Expressivity}\label{appendix:expressivity}
Expressivity typically refers to the capacity of a parameterized quantum circuit to cover the available states of a $n$-qubit system. One way to determine this capacity is to quantify how distant a given ensemble of states $\ket{\psi({\bm{\theta})}} = U(\bm{\theta})\ket{\psi_0}$ is from a $t$-design according to the Hilbert-Schmidt norm~\cite{Sim.2019}. Here, $\ket{\psi_0}$ refers to an initial state, where a parameterized circuit $U(\bm{\theta})$ of parameters $\bm{\theta}$ is applied to.  Formally, this difference is defined as, 
\begin{equation}
\label{eq:expressiblity_definition_appendix}
 \epsilon_{\Theta} = \norm{\underbrace{\int_{\text {Haar }}(|\psi\rangle\langle\psi|)^{\otimes t} d \psi}_{\mathcal{V}_{\mathrm{Haar}}} -\underbrace{\int_{\Theta}  (|\psi(\bm{\theta})\rangle\langle\psi(\bm{\theta})|)^{\otimes t}}_{\mathcal{V}_{\mathbb{U}_\Theta}} d (\bm{\theta})}_{1}\,,
\end{equation}
where the first integral is taken over all pure states over the Haar measure, and the second integral is taken over all states $\ket{\psi_{\bm{\theta}}} = U({\bm{\theta}})\ket{\psi_0}$ spanned by the encoding unitary $U({\bm{\theta}})$, according to an ensemble in consideration, $\mathbb{U}_\Theta = \{U(\bm{\theta}) |\bm {\theta} \in \Theta \}$, with $\Theta$ the space of all parameter vectors. With these definitions, it can be shown that  Eq.~\eqref{eq:expressiblity_definition_appendix}, can be written as
\begin{equation}
\label{eq:expressivity_simplified_appendix}
 \epsilon_{\Theta}^2=  \mathbb{E}_{\boldsymbol{\theta}, \boldsymbol{\phi}}\left[\left(\left|\left\langle\psi_{\boldsymbol{\theta}} \mid \psi_{\boldsymbol{\phi}}\right\rangle\right|^2\right)^t\right] - \frac{(t) !(2^n-1) !}{(t+2^n-1) !}\,.
\end{equation}
In the QK framework, the ensemble under consideration is given by the set of all of the encoded unitaries in our input set $\mathcal{X}$, i.e. $\mathbb{U}_{\mathcal{X}} = \{U(\mathbf{x}) | \mathbf{x} \in \mathcal{X}\}$. Therefore, the parameterized entries of the unitary correspond to the encoded data points and the parameter space is given by the data entries $\mathbf{x}$ and 
$\mathbf{x}^\prime$ of the kernel,
\begin{align}
    \mathbb{E}_{\boldsymbol{\theta}, \bm{\phi}} \left[\left(\left|\left\langle\psi_{\boldsymbol{\theta}} \mid \psi_{\boldsymbol{\phi}}\right\rangle\right|^2\right)^t\right]  &= \left(\mathbb{E}_{\mathbf{x}, \mathbf{x}^\prime}\left[\left(\left|\left\langle\psi_{\mathbf{\tilde{x}}} \mid \psi_{\mathbf{\tilde{x}}^\prime}\right\rangle\right|^2\right)^t\right]\right) \notag \\ 
     &= \left(\mathbb{E}_{\mathbf{x}, \mathbf{x}^\prime}\left[\left( k^{\text{FQK}}(\mathbf{\tilde{x}}, \mathbf{\tilde{x}}^\prime)  \right)^t\right]\right)\,. 
\end{align}
Therefore, it follows
\begin{equation}
 \epsilon_{\mathcal{X}}^2 =  \frac{-(t) !(2^n-1) !}{(t+2^n-1) !} + \mathbb{E}_{\mathbf{x}, \mathbf{x}^\prime}\left( k^{FQ}(\mathbf{\tilde{x}}, \mathbf{\tilde{x}}^\prime)^t\right)\,.
\end{equation}
For the first term, we consider a 2-design, i.e., $t=2$ and for the second term, in the case of a uniform distribution with the QK of Eq.~\eqref{eq:separable_R_X}, we perform the integration
\begin{equation}
 \mathbb{E}_{\mathbf{x}, \mathbf{x}^\prime} \left[ k_{R_X}^{\text{FQK}}(\boldsymbol{\mathbf{\tilde{x}}}, \boldsymbol{\mathbf{\tilde{x}}^\prime})^2 \right] = \int_{\mathbf{\tilde{x}}, \mathbf{\tilde{x}}^\prime} k_{R_X}^{\text{FQK}}(\boldsymbol{\mathbf{\tilde{x}}}, \boldsymbol{\mathbf{\tilde{x}}^\prime})^2 p(\mathbf{\tilde{x}}) p(\mathbf{x^\prime}) \mathrm{d}\mathbf{\tilde{x}} \mathrm{d}\mathbf{\tilde{x}^\prime}\,,
\end{equation}
which yields
\begin{align}\label{eq:expectation_value_main_text}
   \mathbb{E}_{\mathbf{x}, \mathbf{x}^\prime} \left( k_{R_X}^{\text{FQK}}(\boldsymbol{\mathbf{\tilde{x}}}, \boldsymbol{\mathbf{\tilde{x}}^\prime})^2\right) =
   \left( \frac{3}{8} - \frac{\cos(2\pi Lc)}{4\pi^2 (Lc)^2} \right. \nonumber \\
   \left. - \frac{\cos(4\pi Lc)}{64\pi^2 (Lc)^2} + \frac{17}{64\pi^2 (Lc)^2} \right)^n\,.
\end{align}

\subsection{Largest eigenvalue}
\label{appendix:largesteigenvalue}
The eigenvalue spectra of the continuous analogous of $\mathbf{K}_{R_X}^{\text{FQK}}$ (kernel integral operator) for $n_q=1$, was obtained for a uniform distribution $x_i \in \mathcal{X}= [-\pi, \pi ]$ in Ref.~\cite{canatatheorybandwidthenablesgeneralization} as
\begin{align}
\lambda_1 &= \frac{3}{8} + \frac{1}{8} \frac{\sin(2\pi Lc)}{2\pi Lc} \notag \\
 &+ \frac{1}{8} \sqrt{\left(1 - \frac{\sin(2\pi Lc)}{2\pi Lc} \right)^2 + 16 \left(\frac{\sin{(\pi Lc)}}{\pi Lc}\right)^2}\,,\\
\lambda_2 &= \frac{1}{4} - \frac{1}{4} \frac{\sin(2\pi Lc)}{2\pi Lc}\,,\\
\lambda_3 &= \frac{3}{8} + \frac{1}{8} \frac{\sin(2\pi Lc)}{2\pi Lc} \notag\\
&- \frac{1}{8} \sqrt{\left(1 - \frac{\sin(2\pi Lc)}{2\pi Lc} \right)^2 + 16 \left(\frac{\sin{(\pi Lc)}}{\pi Lc}\right)^2}\,,\\
\lambda_4 &= 0\,.
\end{align}
For an $n_q$ qubit system, Ref. \cite{canatatheorybandwidthenablesgeneralization} showed that,
\begin{equation}
\eta_{n_1 n_2 n_3 n_4} = \lambda_1^{n_1} \lambda_2^{n_2} \lambda_3^{n_3} \lambda_4^{n_4}, 
\end{equation}
where $n_1, n_2, n_3, n_4$ are positive integers (including zero), such that $n_q = n_1 + n_2 + n_3 + n_4 $. From here, it can be  seen that the largest eigenvalue $\eta_{\text{max}}$ is given by,
\begin{align}
\label{eq:max_eigenvalue}
\eta_{\text{max}}(c, n_q) &=  \lambda_1^{n} \\
&= \left(\frac{3}{8} + \frac{1}{8} \frac{\sin(2\pi Lc)}{2\pi Lc} \right. \\
&\quad \left. + \frac{1}{8} \sqrt{\left(1 - \frac{\sin(2\pi Lc)}{2\pi Lc} \right)^2 + 16 \left(\frac{\sin{(\pi Lc)}}{\pi Lc}\right)^2}\right)^n \notag
\end{align}

%\clearpage
\onecolumngrid

\section{Supporting figures for bandwidth-tuned kernels}
\label{appendix:figures_for_bandwidth_tuned}
In this section, we extend the findings from Sec.~\ref{sec:results} and report on results for the remaining datasets and data encoding circuits.

\subsection{Overview of Frobenius distances and Geometric differences across datasets}\label{appendix:heatmapsection}

\begin{figure*}[htbp]
    \centering
    \makebox[0.32\linewidth]{(A)}
    \makebox[0.32\linewidth]{(B)}
    \makebox[0.32\linewidth]{(C)}\\
    \includegraphics[width=0.32\linewidth]{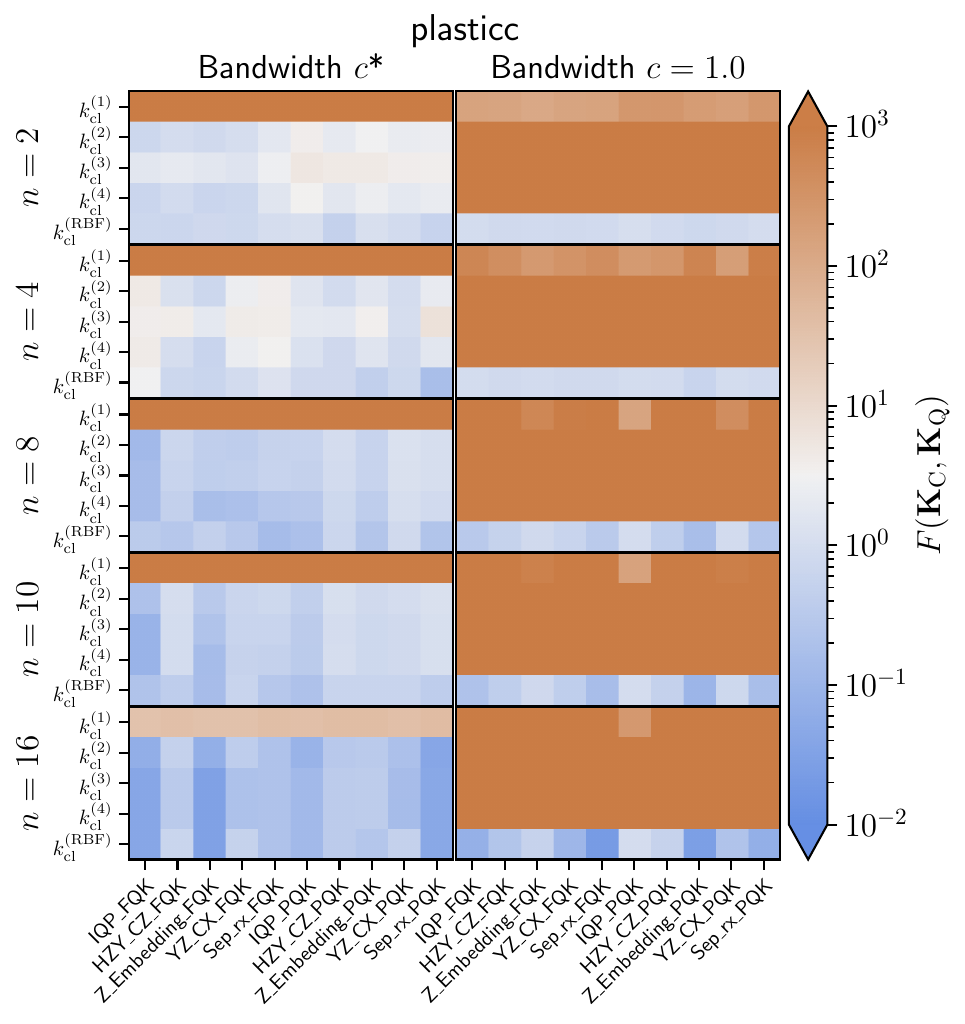}
    \includegraphics[width=0.32\linewidth]{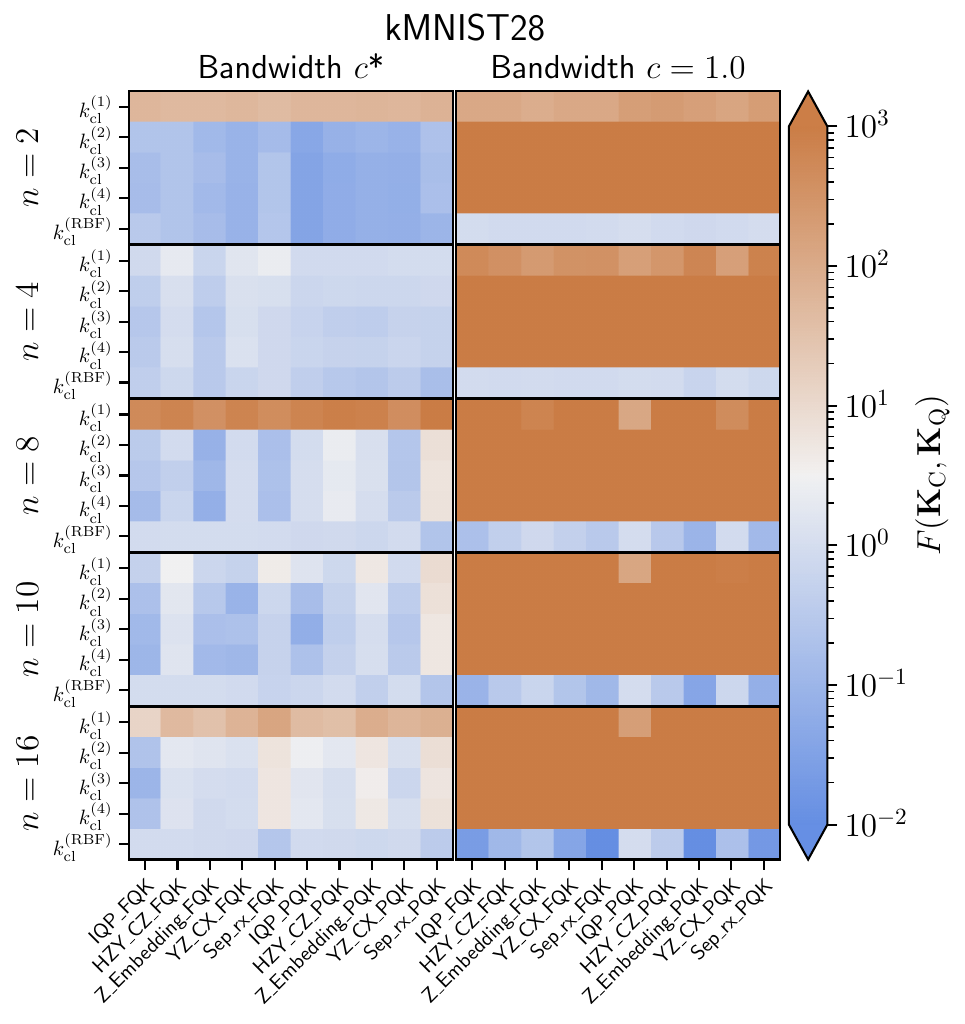}
    \includegraphics[width=0.32\linewidth]{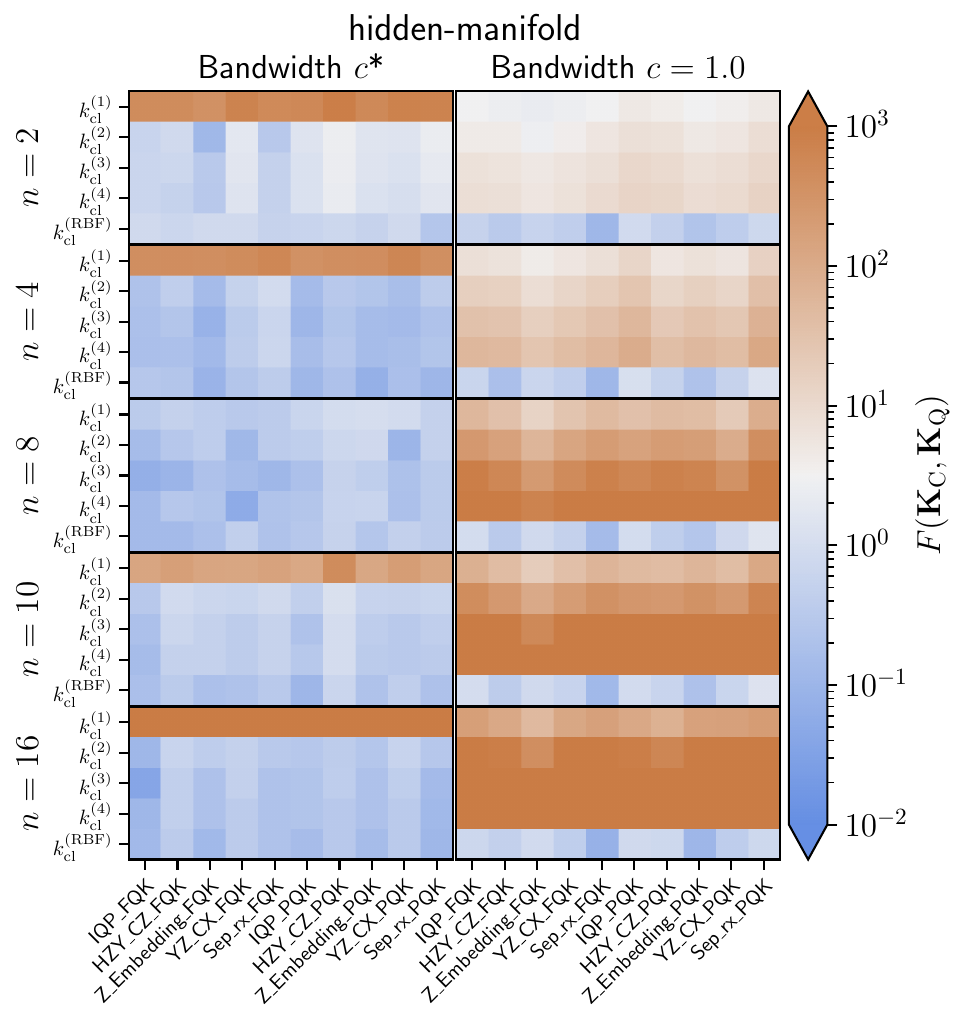}\\
    
    % Second row
    \makebox[0.32\linewidth]{}
    \makebox[0.32\linewidth]{}
    \makebox[0.32\linewidth]{}\\
    \includegraphics[width=0.32\linewidth]{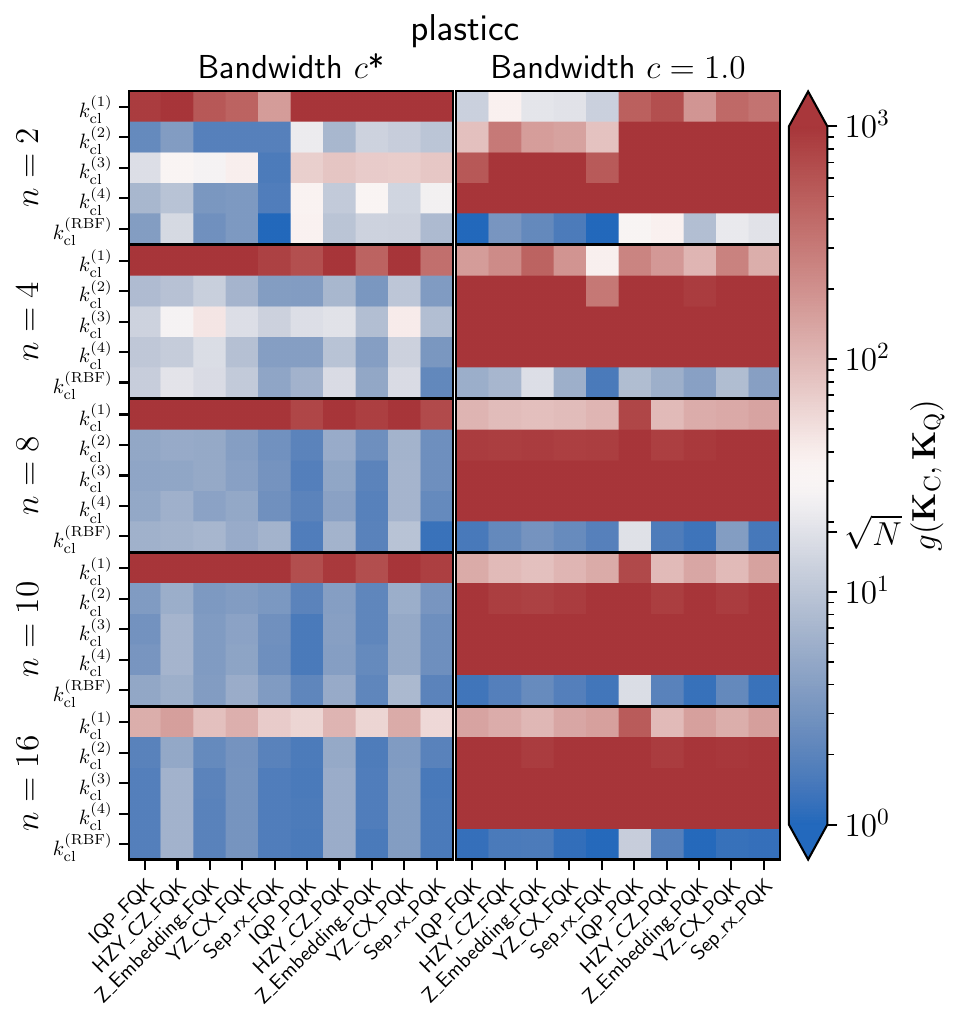}
    \includegraphics[width=0.32\linewidth]{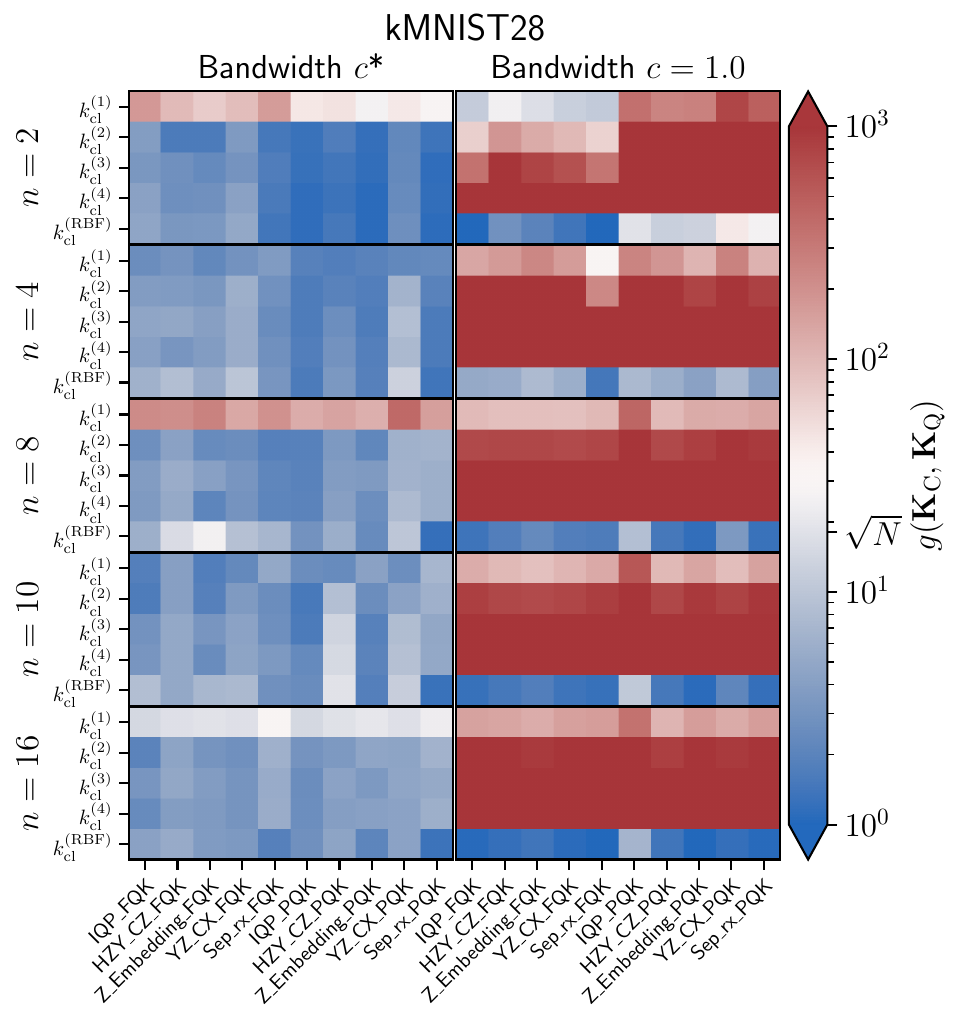}
    \includegraphics[width=0.32\linewidth]{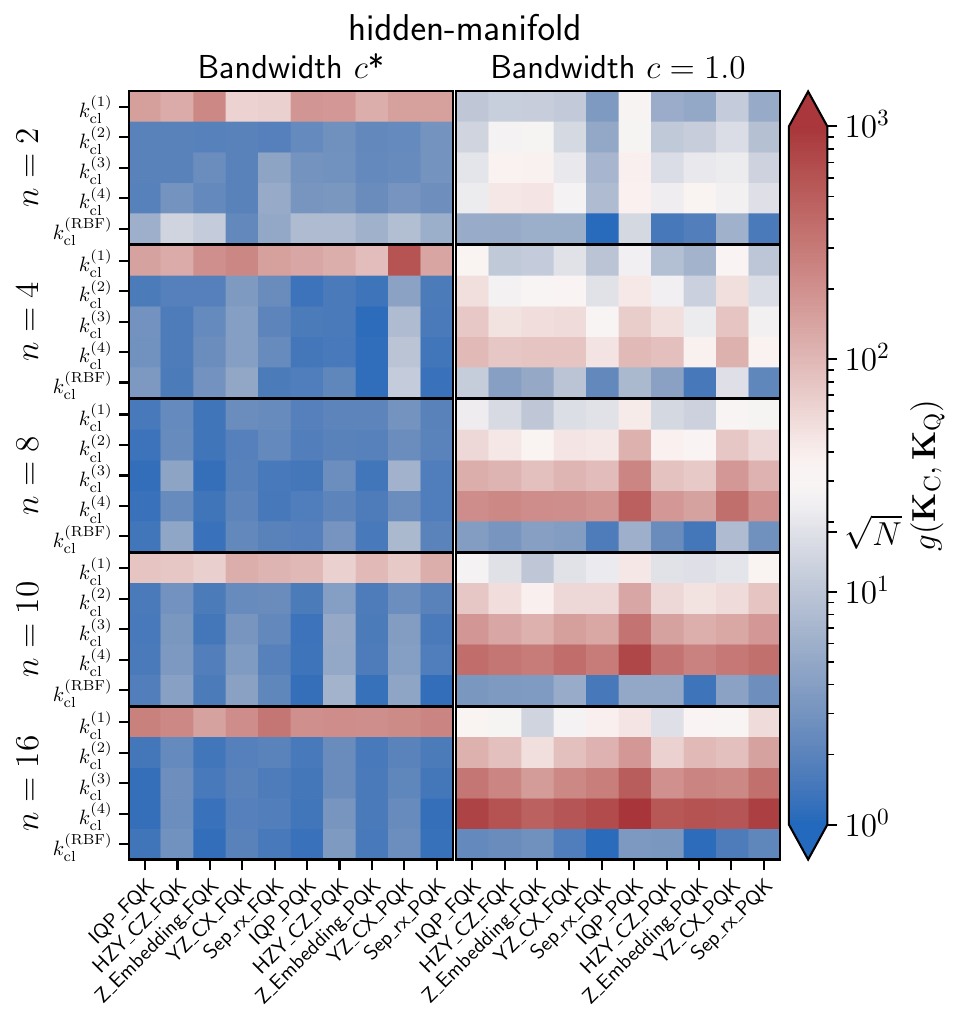}

    \caption{Upper/Lower panel: Frobenius distances/Geometric difference according to Eq.~\eqref{eq:frobenius_distance} / Eq.~\eqref{eq:geo_diff_reg} heatmap between kernel matrices of all the studied datasets. From left to right: \plasticcName, \kmnistName and \, \hiddenmanifoldName. Each block represents the distance between the quantum (on the $x$-axis) and classical (on the $y$-axis) kernel matrices. For each dataset two different bandwidths are shown. The $c^*$ column corresponds to the matrices with optimal bandwidth (highest ROC-AUC score), and the second column $c=1$. }
    \label{fig:heatplotmapsdifferences}
\end{figure*}

The comparisons between quantum and classical models performed in the main text, are limited to $\kpoly{2}$ and $\kRBF$ and only the \plasticcName \, dataset. In Fig.~\ref{fig:heatplotmapsdifferences}, we report the median values of the Frobenius distances (A-C) and the geometric difference (D-F) for all datasets and all data encoding circuits investigated.  Each subfigure contains two columns, the quantities obtained for $c^*$ and for $c=1$ in the left and right columns, respectively. The small values of $c^*$, lead to QKs that can be captured by classical models with $d\geq2$, as seen in the geometric difference plots and mirrored by the Frobenius distances. For $c=1$, RBF kernels successfully capture the CKs. Note that, although the similarity holds across datasets, its efficacy still depends on the underlying dataset. For example, in Figs.~\ref{fig:heatplotmapsdifferences}(B) and \ref{fig:heatplotmapsdifferences}(C), we see larger Frobenius distances for \hiddenmanifoldName\, than for \kmnistName\,.

\subsection{Quantities as a function of bandwidth}\label{appendix:quantities_vs_bandwidth}

\begin{figure*}[htbp]
    \centering
    \includegraphics[width=0.2526\linewidth]{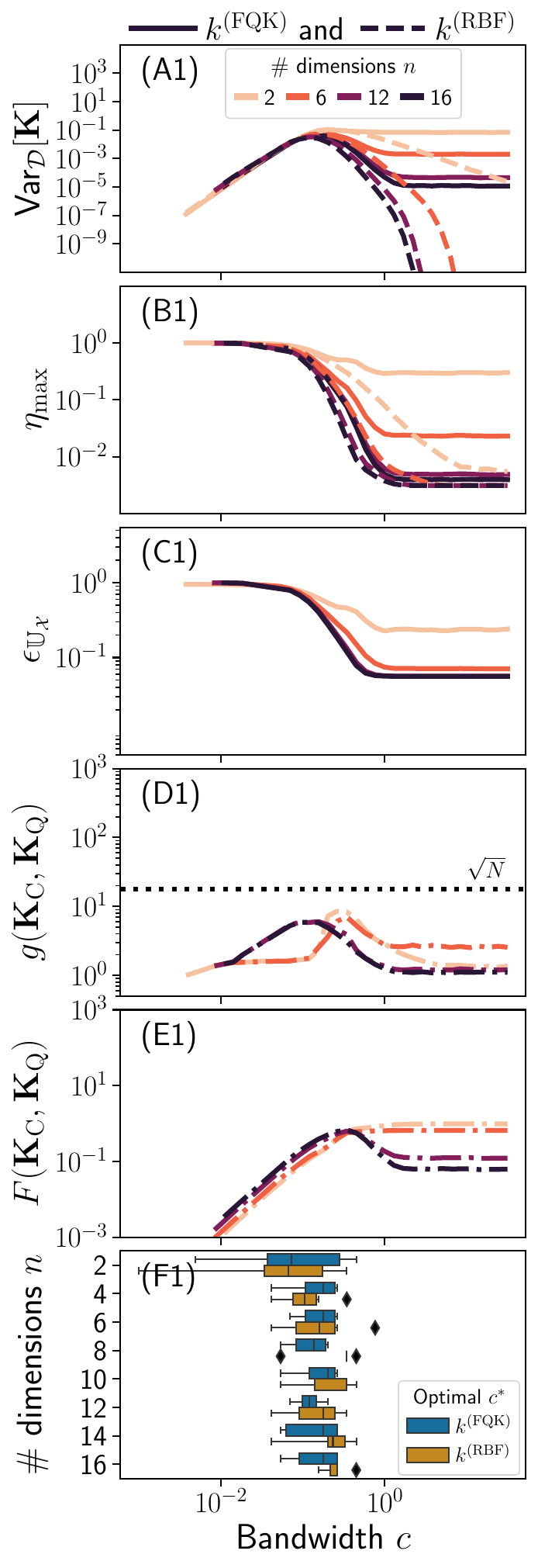}
    \includegraphics[width=0.205\linewidth]{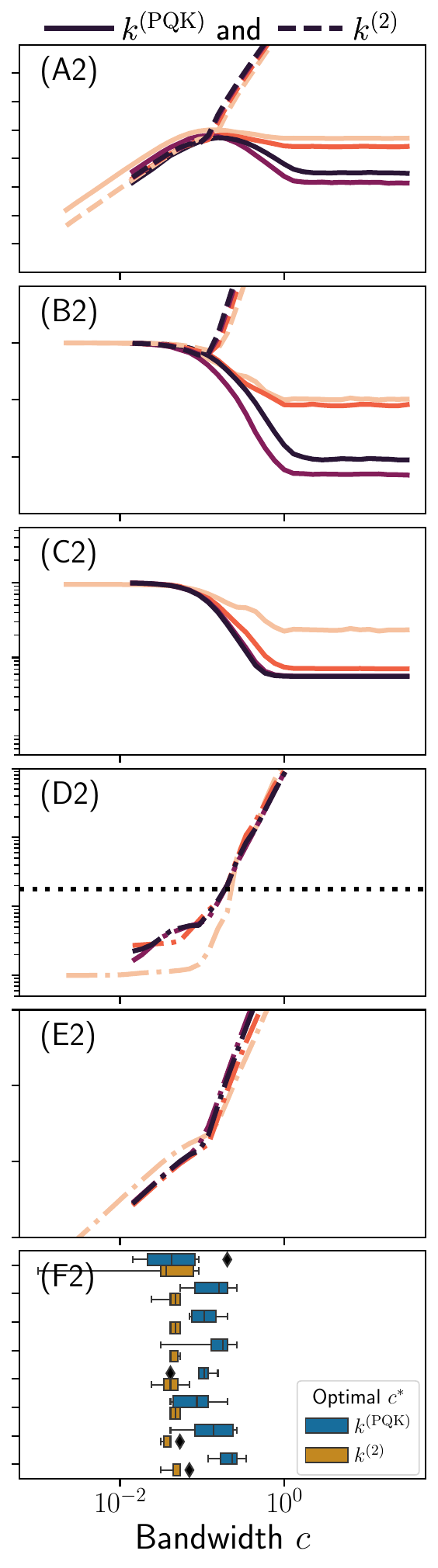}
    \includegraphics[width=0.205\linewidth]{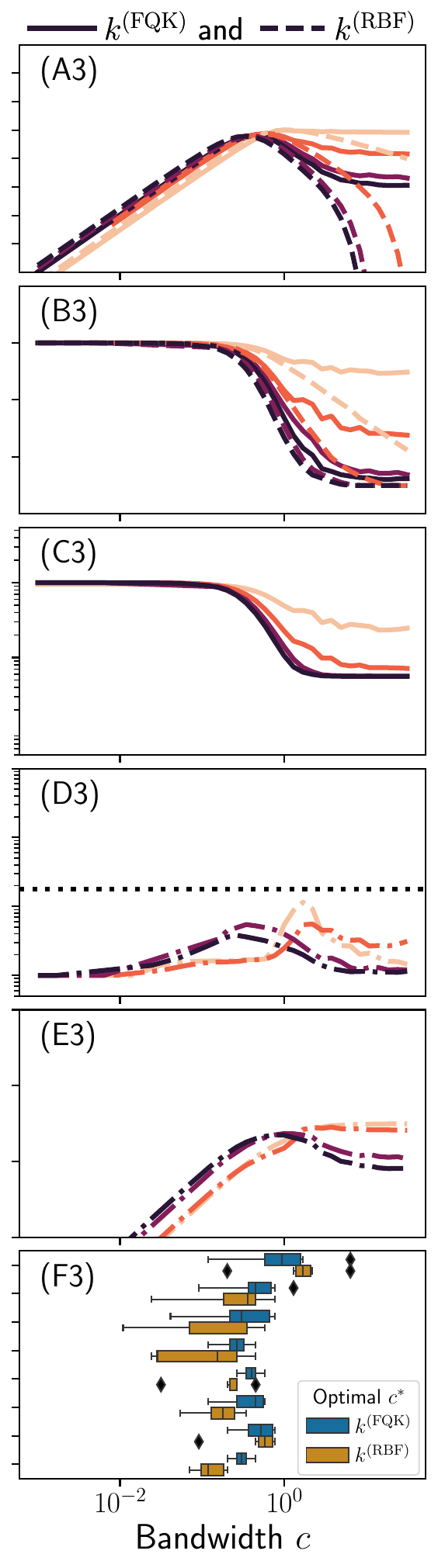}
    \includegraphics[width=0.205\linewidth]{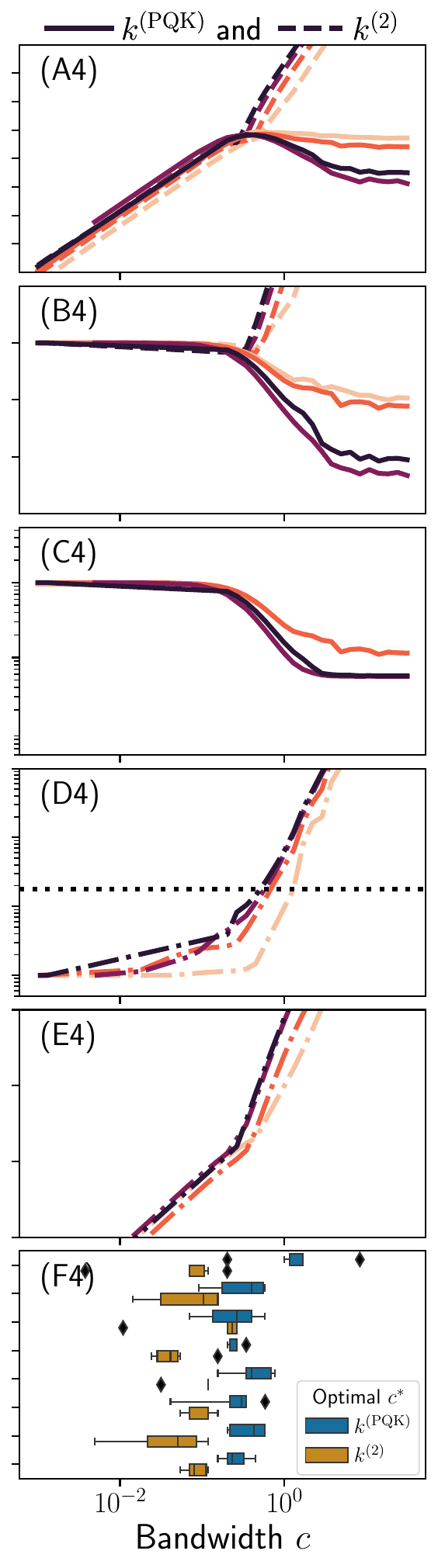}
    \caption{Panels 1-2 (3-4) extend Fig. \ref{fig:metrics_vs_bandwidth} for \kmnistName \, (\hiddenmanifoldName) dataset. }
    \label{fig:QK_vs_c_appendix}
\end{figure*}

Furthermore, in Sec.~\ref{sec:models_vs_c} we discuss how key quantities of the CKs and QKs change as a function of the bandwidth parameter for the \plasticcName \, dataset. Here, we extend the figures for the remaining datasets. In Figs.~\ref{fig:QK_vs_c_appendix}(1-2), we show the equivalent results for the \kmnistName \, dataset and in Figs.~\ref{fig:QK_vs_c_appendix}(3-4) for \hiddenmanifoldName \,. The same three regimes are observed for all datasets. For completeness, the simulations with the other encoding circuits are available in Ref.~\cite{robertogit}. 

\subsection{Quantities as a function of number of dimensions}\label{appendix:quantities_vs_n}
\begin{figure*}[h]
    \centering
    \includegraphics[width=1\linewidth]{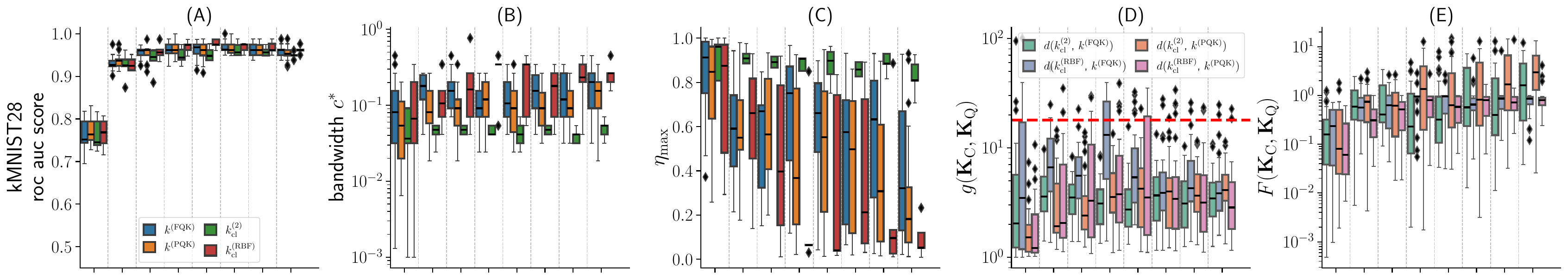}
    \includegraphics[width=1\linewidth]{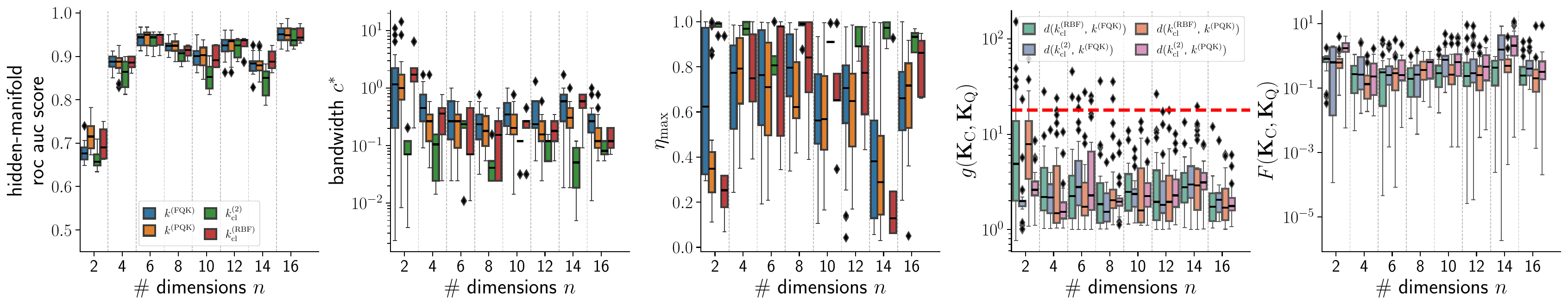}
    \caption{Plots similar to Fig. \ref{fig:quantities_vs_n} for the remaining datasets. First and second row, for the \kmnistName \,, and \hiddenmanifoldName \, datasets respectively.}
    \label{fig:quantities_vs_n_appendix}
\end{figure*}

Similar to Sec.~\ref{sec:models_vs_n} and Fig.~\ref{fig:quantities_vs_n}, we show the aggregated results for the \kmnistName \, and \hiddenmanifoldName \, datasets in Fig.~\ref{fig:quantities_vs_n_appendix}. As in the main text, we observe that optimal RBF kernels and optimal QKs resemble each other, with similar largest eigenvalues and small geometric difference. The Frobenius distance follows the geometric difference. Furthermore, if the optimal bandwidths are small, the truncated RBF kernels capture the QKs.

\clearpage
\subsection{Eigenvalues \label{appendix:eigenvalues}}

\begin{figure*}[h]
    \centering
    \includegraphics[width=\linewidth]{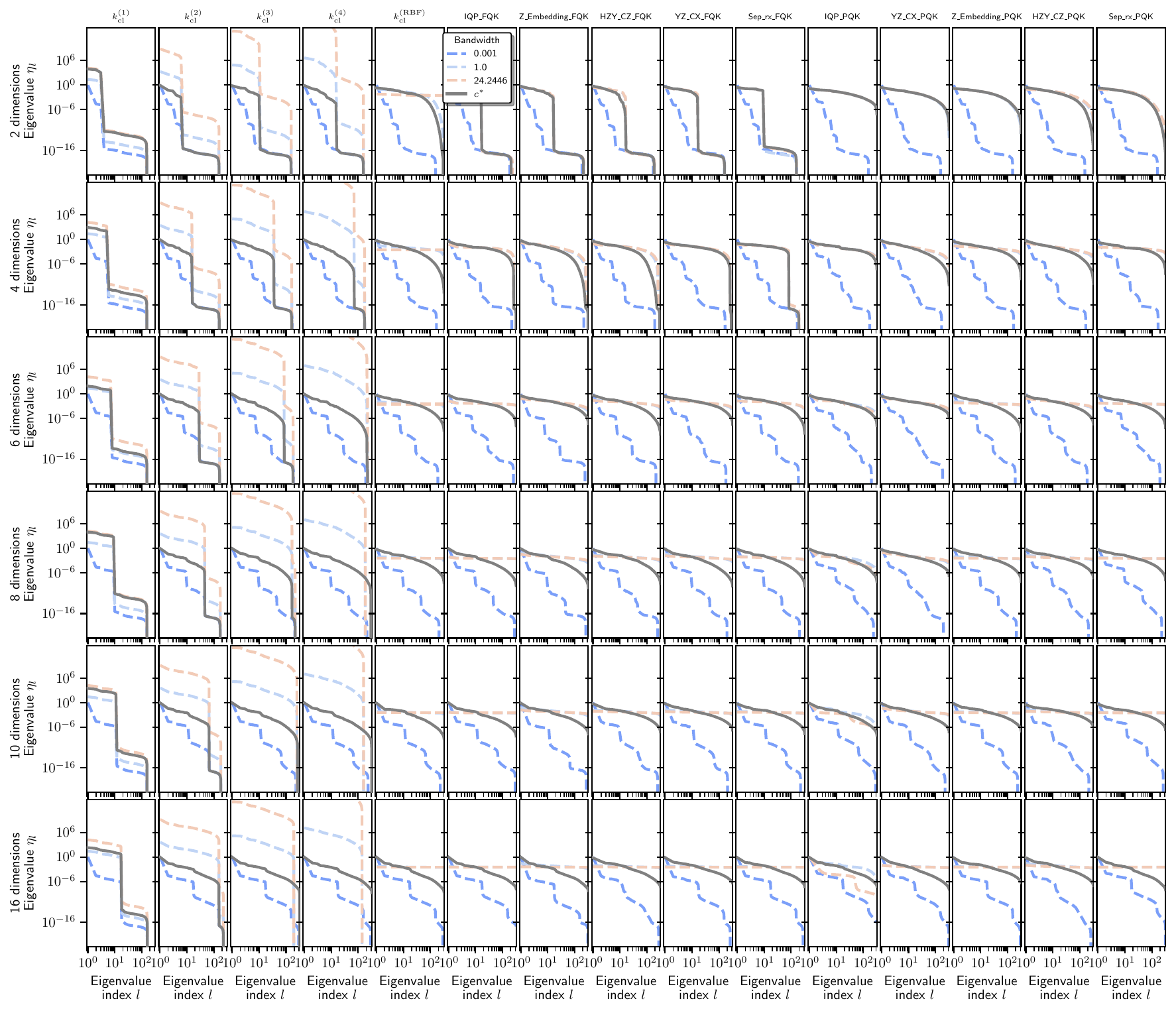}
    \caption{\plasticcName \, dataset. Eigenvalue spectra for increasing dimensionality of classical and quantum models used for the experiments. \textbf{From left to right}, different kernel models: $k^{(1)}$, $k^{(2)}$, $k^{(3)}$, $k^{(4)}$, RBF, IQP, Z embedding, HZY\_CZ, YZ\_CX and Separable $R_X$. \textbf{From top to bottom}, different dimensions: 2, 4, 8, 10, 16. }
    \label{fig:eigenvalues_appendix_plasticc}
\end{figure*}

We extend the investigation of the largest eigenvalue of Fig.~\ref{fig:quantities_vs_n}~(B) to present the full eigenvalue spectra for most of the simulated dimensions and all classical and quantum models for the \plasticcName \, dataset in Fig.~\ref{fig:eigenvalues_appendix_plasticc}. The other simulations for \kmnistName \, and \plasticcName \, are available on GitHub~\cite{robertogit}.
In these plots, it becomes clear that the similarity of the QKs to the polynomial kernels is due to the small size of the bandwidth. At the same time, the resemblance between the RBF kernels and QKs is evident across all bandwidth values.

\clearpage

\section{Dependence of the three regime framework with the number of layers and number of data points }

As discussed in Secs.~\ref{sec:models_vs_c} and~\ref{sec:analytical_toy_model}, the three regimes observed therein, are also affected by the number of re-uploading layers and the number of training points. Here, we characterize the effect of these parameters.

\subsection{Effect of finite number of data points in toy model}
\label{appendix:effect_of_finite_number}

\begin{figure*}[h]
    \centering
    \includegraphics[width=\linewidth]{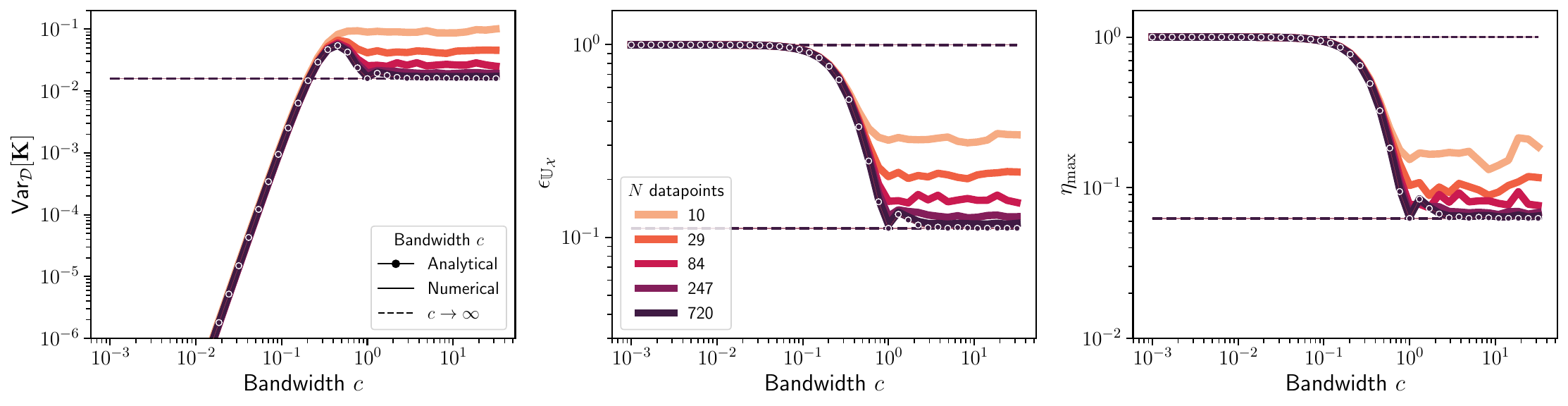}
    \caption{Effect of the number of datapoints in the key quantities of a kernel matrix for $n=4$ qubits. As the number of points increases, the numerical values agree with the analytical prediction and with the $c \rightarrow \infty$ limit.}
    \label{fig:analyticalmodeldatapoints_nfixed}
\end{figure*}

\begin{figure*}[h]
    \centering
    \includegraphics[width=\linewidth]{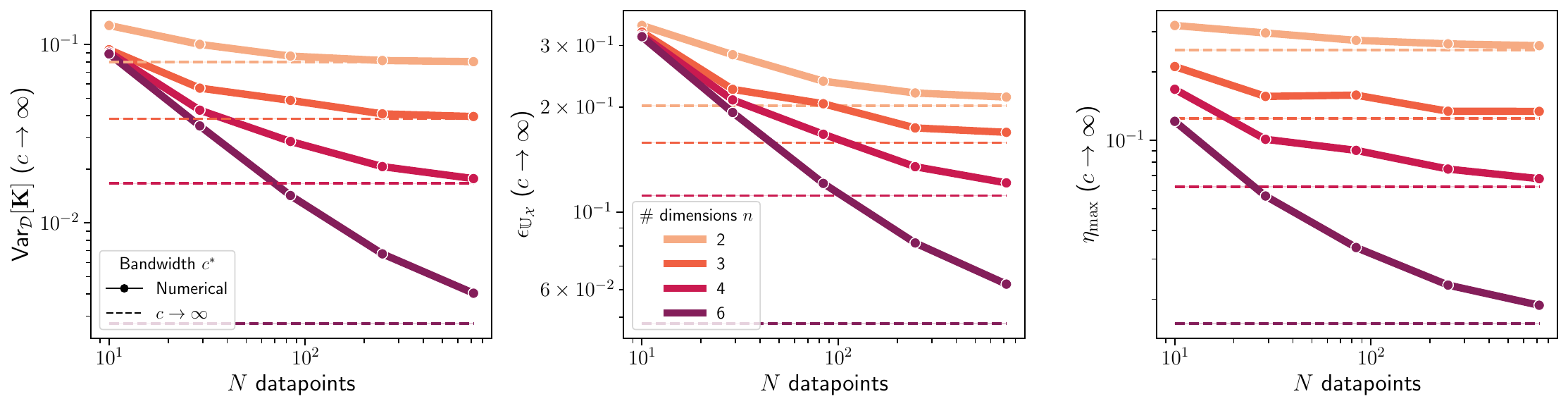}
    \caption{Effect of the number of data points in the key quantities for the $c\rightarrow\infty$ range for different number of dimensions. As the number of points increases, the numerical values agree with the analytical prediction and with the $c \rightarrow \infty$ limit.}
    \label{fig:analyticalmodeldatapoints_nchanging}
\end{figure*}

The derivation of the analytical equations in Sec.~\ref{sec:analytical_toy_model}, takes into account sampling from a probability distribution $p(\mathbf{x})$. On the other hand, the numerical simulations only have access to a finite number of data points. In order to match the analytical and numerical equations, the number of sampled points has to be sufficiently large.  In Fig.~\ref{fig:analyticalmodeldatapoints_nfixed}, we reproduce the plots of Sec.~\ref{sec:analytical_toy_model} but we fix the dimensionality $n=4$ and increase the number of data points. This is insightful, as the only regime which is largely impacted by the number of datapoints is the large bandwidth range. The overall function behavior is still the same and for large $c$ the quantities plateaus, however, the plateau agrees with the analytical limit only for $N$ sufficiently large. Moreover, as the dimensionality increases, saturating the limit becomes harder (Fig.~\ref{fig:analyticalmodeldatapoints_nchanging}) as the Hilbert space increases exponentially. Thus, providing an intuition about why Fig.~\ref{fig:metrics_vs_bandwidth} do not endlessly decrease with dimensionality.

\subsection{Effect of number of layers in the three regime framework}
\label{appendix:effect_of_layers}

\begin{figure*}[h]
    \centering
    \makebox[0.271\linewidth]{(A)}
    \makebox[0.23\linewidth]{(B)}
    \makebox[0.23\linewidth]{(C)}
    \makebox[0.23\linewidth]{(D)}
    \\\includegraphics[width=0.271\linewidth]{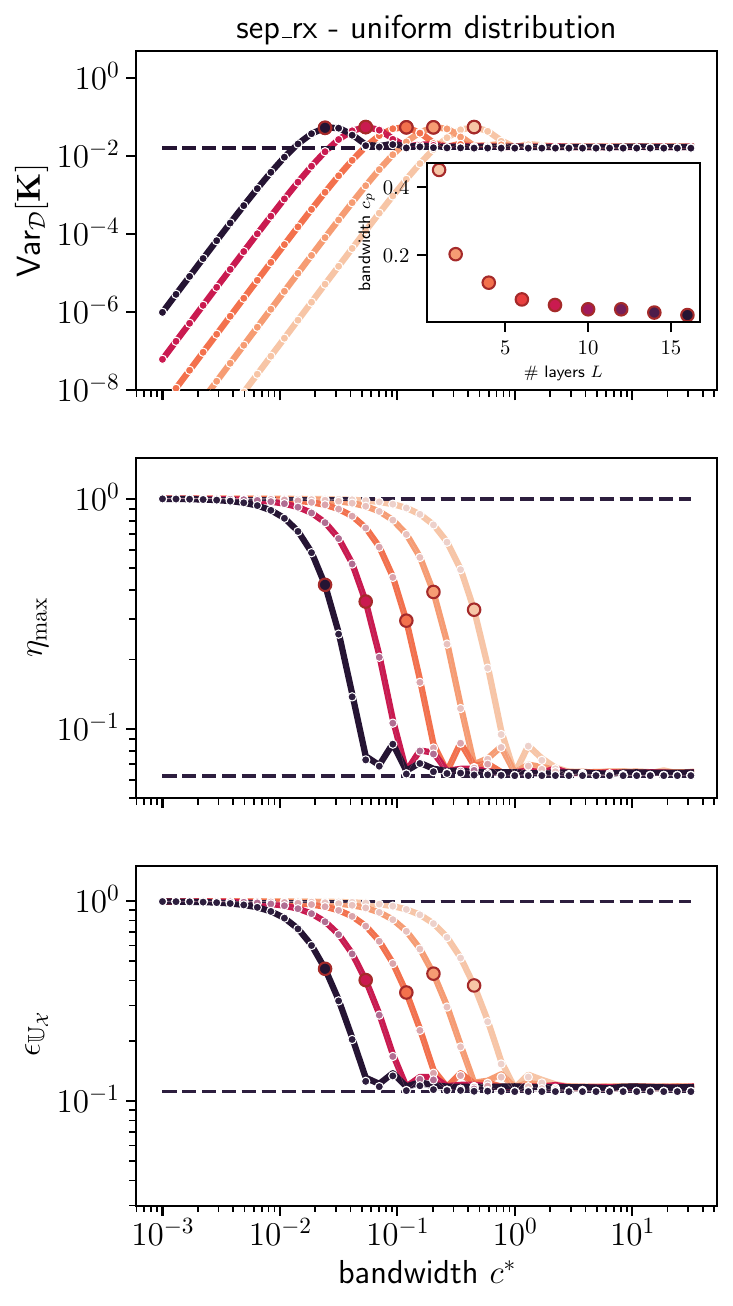}
    \includegraphics[width=0.23\linewidth]{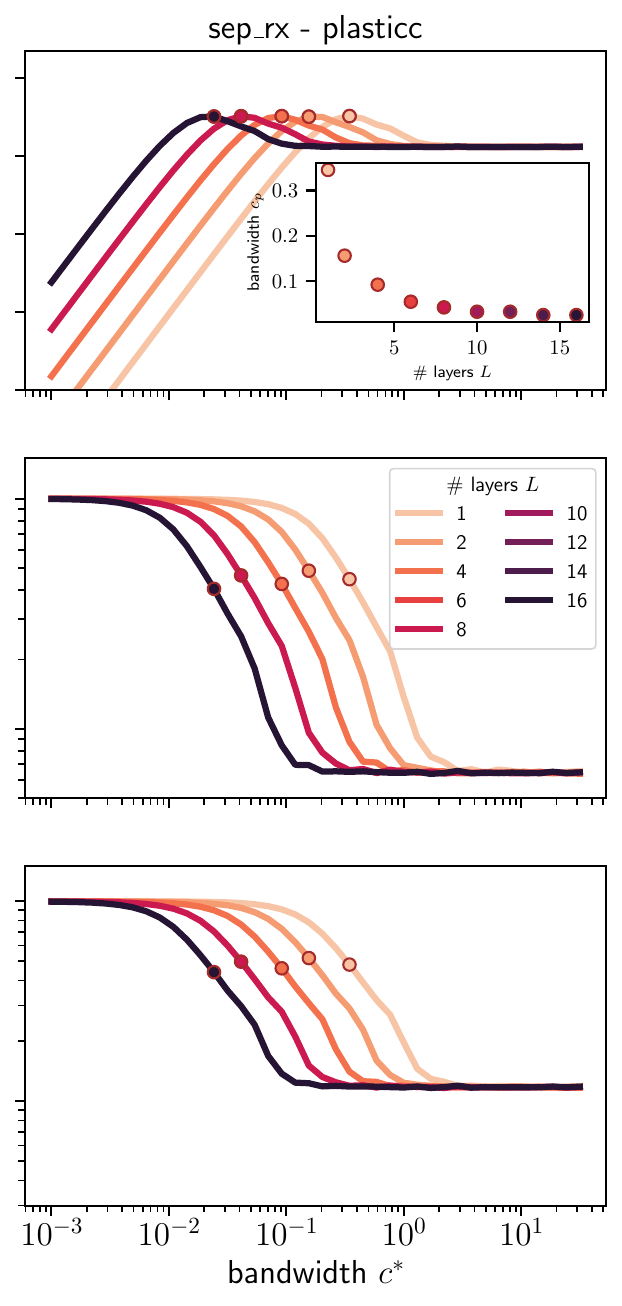}        \includegraphics[width=0.23\linewidth]{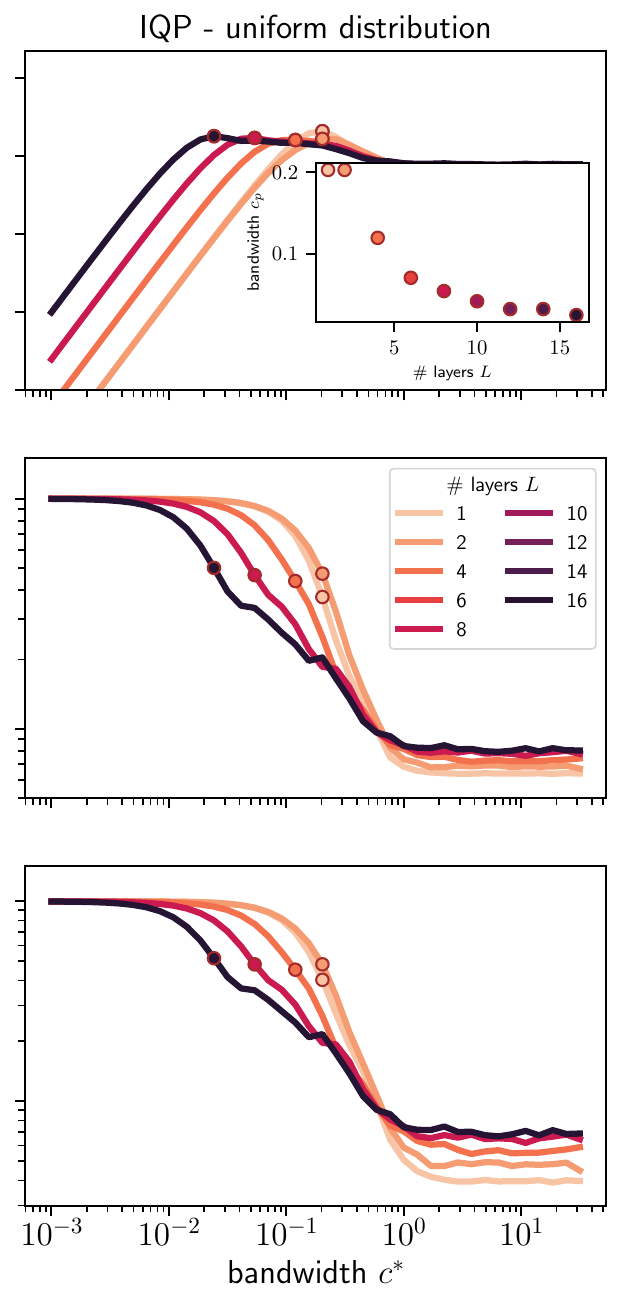}
    \includegraphics[width=0.23\linewidth]{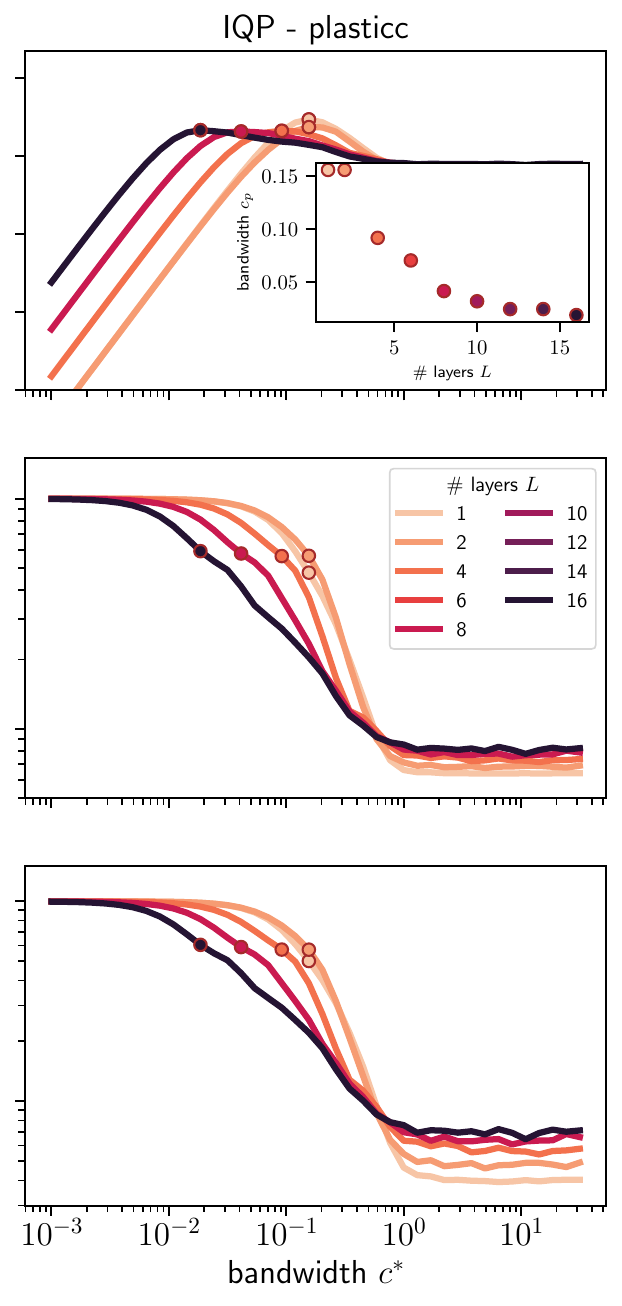}
    \caption{The regimes described in the main text are shifted to the left with increasing number of reuploading layers for trivial and nontrivial distributions.
    On columns 1 and 2 (3 and 4), Separable $R_x$ and IQP encoding circuits \cite{Hubregtsen.2022} for a uniform distribution (\plasticcName \, distribution). The first column contains also the analytical predictions of Eq. \ref{eq:toymodelmainuniform} in dotted white points. The insets in the first row, corresponds to changes in the peak bandwidth values (solid points)  of the variance $c_p$, as a function of the number of layers. For these simulations we used $N=720$ and $n=4$. The lines contain simulations for $L \in \{1,2,4,8,16\}$, whereas the inset also includes $L \in \{6,10,12,14\}$. }
    \label{fig:effect_of_layers}
\end{figure*}

Increasing the number of re-uploading layers affects the QKs. In the case of $k_{R_X}$, cf. Eq.~\eqref{eq:separable_R_X} with a uniform distribution, it can be shown that this leads to an extra rescaling of the angles, cf. Eq. \eqref{eq:toymodelmainuniform}, which reflects in a general shift to smaller bandwidths of the three regimes. This can be observed in Fig.~\ref{fig:effect_of_layers}~(A). The analytical equation accurately captures the displacement. The shift is also observed in other distributions, as can be seen for the same QK in the \plasticcName \, dataset in Fig.~\ref{fig:effect_of_layers}~(B). Finally, we observe that for other encoding circuits, the shift appears as well but shows some other effects. In the case of the circuit from Ref.~\cite{Hubregtsen.2022}, although the variance peaks displaces, the entire functions does not, cf., Figs.~\ref{fig:effect_of_layers}~(C,D). 

\clearpage
\end{document}